\documentclass[preprint,12pt,authoryear]{elsarticle}
\usepackage{amssymb}
\usepackage{amsmath}
\usepackage[top=0.9in, bottom=0.9in, left=0.9in, right=0.9in]{geometry}
\usepackage{amssymb,booktabs,subfig,xcolor,microtype,setspace,bm,paralist,dsfont,longtable} 
\usepackage{natbib}
\usepackage{url}
\usepackage[pdftex,colorlinks=true,hypertexnames=false]{hyperref}
\definecolor{darkblue}{rgb}{0,0,.6}
\hypersetup{citecolor=darkblue,linkcolor=darkblue,urlcolor=darkblue}
\usepackage{amsfonts}
\usepackage{epsfig}
\usepackage{graphics}
\usepackage{palatino,eulervm}
\usepackage{graphicx}
\usepackage{lscape}
\usepackage{rotating}
\usepackage[linewidth=1pt]{mdframed}
\usepackage{orcidlink}
\usepackage{subcaption}
\usepackage{multirow}
\usepackage{afterpage}
\allowdisplaybreaks[4]

\usepackage[normalem]{ulem}

\providecommand{\U}[1]{\protect\rule{.1in}{.1in}}

\graphicspath{{plots/}}

\usepackage{amsthm,thmtools}
\usepackage{mathrsfs}

\makeatletter
\def\th@newremark{\th@remark\thm@headfont{\bfseries}}
\makeatletter
\theoremstyle{newremark}
\newtheorem{remark}{Remark}
\newtheorem{proposition}{Proposition}

\declaretheoremstyle[
  spaceabove=6pt, spacebelow=6pt,
  headfont=\bfseries,
  notefont=\mdseries, notebraces={(}{)},
bodyfont=\normalfont,
  postheadspace=0.5em]{mystyle}

\newsavebox\CBox
\def\textBF#1{\sbox\CBox{#1}\resizebox{\wd\CBox}{\ht\CBox}{\textbf{#1}}}
\newcommand{\commHS}[1]{{\leavevmode\color{purple}#1}}

\newcommand{\Rlogo}{\protect\includegraphics[height=1.8ex,keepaspectratio]{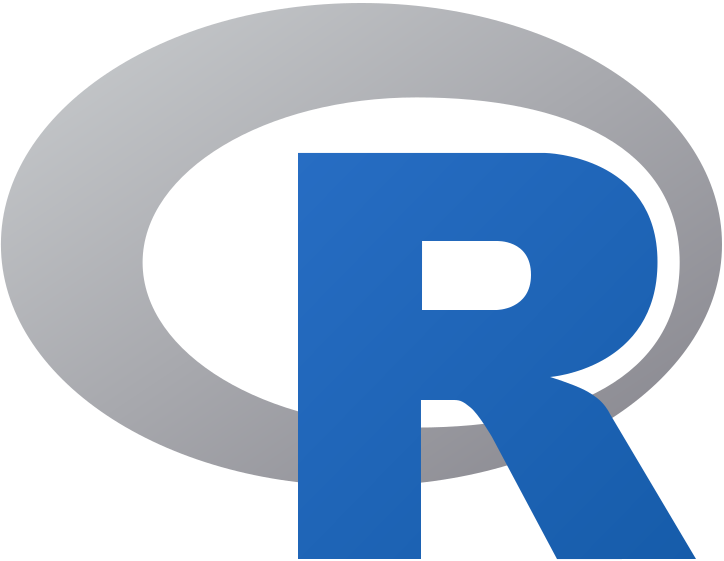}}

\journal{International Journal of Forecasting}

\begin{document}

\begin{frontmatter}

\title{Enhancing Mortality Forecasting with Ensemble Learning: \mbox{A Shapley-Based Approach}}

\author{Giovanna Bimonte\orcidlink{0000-0003-0205-2704}\qquad Maria Russolillo\orcidlink{0000-0001-9455-7085}}

\affiliation{organization={Department of Economics and Statistics, University of Salerno},
            addressline={Via Giovanni Paolo II, 132}, 
            city={Fisciano},
            postcode={84084}, 
            country={Italy}}


            \author{Yang Yang\orcidlink{}} 
\affiliation{organization={School of Information and Physical Sciences, University of Newcastle},
            addressline={University Drive}, 
            city={Callaghan},
            state={NSW},
            postcode={2308}, 
            country={Australia}}

            \author{Han Lin Shang\orcidlink{0000-0003-1769-6430}}
\affiliation{organization={Department of Actuarial Studies and Business Analytics, Macquarie University},
            addressline={4 Eastern Rd}, 
            city={Sydney},
            state={NSW},
            postcode={2109}, 
            country={Australia}}

\begin{abstract}
A well-established insight in mortality forecasting is that combining predictions from a set of models improves accuracy compared to relying on a single best model. This paper proposes a novel ensemble approach based on Shapley values, a game-theoretic measure of each model’s marginal contribution to the forecast. We further compute these SHapley Additive exPlanations (SHAP)-based weights age-by-age, thereby capturing the specific contribution of each model at each age. In addition, we introduce a threshold mechanism that excludes models with negligible contributions, effectively reducing the forecast variance. Using data from 24 OECD countries, we demonstrate that our SHAP ensemble enhances out-of-sample forecasting performance, especially at longer horizons. By leveraging the complementary strengths of different mortality models and filtering out those that add little predictive power, our approach offers a robust and interpretable solution for improving mortality forecasts.
\end{abstract}

\end{frontmatter}


\setstretch{1.5}
\section{Introduction}\label{sec:1}

Mortality forecasting is crucial across several fields, including social security planning, insurance pricing, and public health policymaking. Over the years, the actuarial and demographic literature has proposed numerous stochastic mortality models to capture the dynamics of mortality rates within single populations. Although individual models often demonstrate strengths in specific scenarios, none consistently outperforms others across all datasets or forecast horizons \citep{Makridakis2020}. To address this limitation, many studies have suggested combining forecasts from multiple models \citep[see, e.g.,][]{bates1969combination, Li2005, Shang2011, samuels2017model}, as ensemble methods often yield more accurate predictions than relying on a single "best" model. However, not all model combinations yield better results, as various factors may explain why some combinations are more successful than others. Understanding the factors that underlie the superior performance of certain ensembles remains a critical challenge, and it has been extensively studied to determine optimal model combinations.

An effective ensemble strategy should ideally account for each model's relative contributions and dynamically adjust their weights based on predictive value. One common approach to building ensembles involves assigning equal weights to all models, effectively averaging their forecasts. Although simple and easy to implement, this method has been shown to outperform individual models in many cases. However, this approach does not account for the inherent differences in model performance across datasets and time periods, which can lead to suboptimal results. From the viewpoint of the bias–variance trade-off, forecast combinations with equal weights tend to be biased but have low variance, whereas combinations with estimated weights are generally unbiased but exhibit high variance.

We introduce a novel ensemble approach to determining the optimal weights for each mortality model using SHAP values \citep{NIPS2017SHAP}, a game-theoretic measure that explains the output of any machine learning model. The SHAP value connects optimal credit allocation with local explanations using the classic Shapley values from game theory, and allows us to allocate each model's contribution to the ensemble based on its marginal performance. The use of the SHAP value in the forecast combination has already been considered in the forecasting community. For example, \cite{FZW24} propose a new method to combine forecasts by estimating weights via Shapley values following a Mincer-Zarnowitz regression. \cite{QZL+22} use the Shapley value in the forecast combination of city gas demand in China. While the estimated weights are often static, the predictive ability of the models varies over time. Thus, it is paramount to consider adaptive weights. For example, \cite{ElliottTimmermann2005} propose a new forecast combination method that lets the combination weights be driven by regime switching in a latent state variable. Our method adapts to the models' varying predictive abilities, improving the reliability and precision of mortality forecasts.

Drawing on recent literature, we focus on two important issues that arise when combining prediction methods: accuracy and diversity \citep{lichtendahl2020some}. The inclusion of poor-performing methods in an ensemble is widely recognized as detrimental to forecast accuracy. At the same time, even well-performing methods may fail to deliver strong results if they lack sufficient diversity. To address both issues, we propose an approach that combines 15 distinct mortality prediction models drawn from the most popular time series extrapolative methods employed by statisticians and actuaries in recent decades. We then refine this ensemble by examining two strategies: one that uses SHAP values without any filtering threshold, and one that applies a threshold to exclude models with lower contributions. The latter procedure preserves diversity within the ensemble and improves point and interval forecast accuracies, highlighting the value of strategic model selection in mortality forecasting. 

The paper is structured as follows. In Section~\ref{sec:2}, we review different mortality models and several ensemble approaches (model equations and constraints are provided in \ref{sec:Appendix_B}). In Section~\ref{sec:3}, we describe the theoretical foundation and implementation of our SHAP weighting scheme. In Section~\ref{sec:4}, we introduce details of implementing the proposed and benchmark methods in the empirical analysis. In Section~\ref{sec:5}, we evaluate the proposed method using extensive mortality data (24 OECD Countries) and compare it with traditional ensemble approaches. Section~\ref{sec:6} discusses ensemble prediction intervals constructed by SHAP-based and traditional methods. Conclusions are presented in Section~\ref{sec:7}, along with some ideas on how the methodology presented here can be extended further. From the bias-variance trade-off aspect, we examine the mean square error decomposition of various weighting schemes in~\ref{sec:app}.

\section{Literature Review}\label{sec:2}

\subsection{Notation}\label{sec:2.0}

We introduce the notation and key elements underlying each selected mortality model, clarifying how different parameter structures capture the variability in mortality rates over age, time, and cohort. Specifically:
\begin{itemize}
\item $m_{x,t}$ is the central mortality rate at age $x$ and in calendar year $t$.
\item $D_{x,t}$ denotes the observed number of deaths, assumed to follow a Poisson distribution.
\item $E_{x,t}$ is the central exposure at age $x$ and year $t$.
\item $\alpha(x)$ represents the baseline level of logarithmic mortality throughout the fitting period, capturing the average age structure.
\item $\kappa_{t}^{(i)}$ are the latent time factors that evolve by calendar year $t$.
\item $\beta^{(i)}(x)$ modulates how each $\kappa_{t}^{(i)}$ affects log-mortality at age $x$.
\item $\gamma_{t-x}$ (or $\gamma_{c}$ with $c = t - x$) captures cohort-specific variation (i.e., by birth year).
\item Summation constraints (e.g., $\sum_{t} \kappa_{t}^{(i)} = 0$ or $\sum_{x} \beta^{(i)}(x) = 1$) remove redundant shift or scaling in model parameters.
\end{itemize}

\subsection{Individual Age-Specific Mortality Forecasting Models}\label{sec:2.1}

Our approach begins by integrating a broad selection of individual mortality models, each capturing different facets of mortality rate dynamics -- such as trends, mortality volatility, linearity, and cohort effects. These models serve as the critical foundation for evaluating our proposed ensemble framework. Building on \cite{SH18}, we focus on a well-established family of discrete-time mortality models (summarized in Tables~\ref{tab:models_1}--~\ref{tab:models_4} in~\ref{sec:Appendix_B}) and combine 15 of the most popular extrapolative forecasting methods used in the recent literature. Although we acknowledge the potential to incorporate a broader set of models into our analysis, we also recognize that implementing a large number of models poses significant computational challenges. Consequently, this curated selection strikes a balance between breadth and tractability, enabling us to assess the performance benefits of our ensemble methodology rigorously.

Based on this framework, Tables~\ref{tab:models_1} and~\ref{tab:models_2} highlight two principal families of models, namely Renshaw-Haberman (RH, models~1--3) and Cairns-Blake-Dowd (CBD, models~4--8), which have been widely applied to capture varying patterns of mortality, including linear trends and age-specific mortality improvements. We adopt the Poisson log-bilinear formulation for mortality tables and its cohort-augmented extension within the Renshaw–Haberman family, linking to the age–period–cohort tradition \citep{BrouhnsDenuitVermunt2002, RenshawHaberman2006, Holford1983}. Within the Renshaw--Haberman family, we retain the basic Poisson Lee--Carter model M1 and its cohort-augmented extensions M2 and M3, which already provide a range of age--period--cohort structures. More general multi-factor RH specifications, such as adding a second singular vector in the spirit of the age-specific enhancement model of \cite{RenshawHaberman2003}, are in principle possible and represent a natural development of this family. However, we do not pursue these richer RH variants here to avoid inflating the number of very closely related specifications relative to the other model families and to keep the overall model universe tractable.

For the Cairns–Blake–Dowd family, we use the baseline two-factor CBD model and the widely applied cohort and quadratic-age extensions (M6–M8), together with the Plat specification as a flexible APC variant \citep{CairnsBlakeDowd2006, CairnsEtAl2009, Plat2009}. The original Cairns--Blake--Dowd (CBD) family of models was primarily developed for old-age mortality, typically from about age 55 upwards, where a parsimonious linear or quadratic specification in age provides a good description of the log-mortality profile. In our empirical setting, however, all 15 models are fitted to the same age range, 0--100, to ensure a homogeneous set of inputs for the ensemble and to allow a fully comparable evaluation of forecast performance. For the CBD-type specifications (M4--M6), this should therefore be understood as an extrapolation of models originally designed for older ages. We do not interpret the CBD parameters at young ages in a structural way; rather, these models are used as additional base learners, whose contributions to the final forecasts are determined ex post by their out-of-sample performance via the SHAP weighting scheme.

In Table~\ref{tab:models_3}, the LC (models~9-12) families adopt parametric structures in modeling mortality rates. In particular, the Booth-Maindonald-Smith (BMS) model represents an enhanced version of the LC model developed to improve the stability and reliability of mortality forecasts. The LC model with life expectancy adjustment introduces modifications to account for variations in life expectancy, thereby improving its ability to capture long-term trends. Conversely, the LC model without score adjustment eliminates the ex-post calibration of parameters, providing a more direct estimation process without additional adjustments. For the Lee–Carter variants, we follow the original singular value decomposition formulation, the calibration refinements of Booth–Maindonald–Smith, and the level-adjustment procedure of Lee–Miller; operational choices around score adjustment are discussed in recent reviews \citep{LeeCarter1992, BoothMaindonaldSmith2002, LeeMiller2001, BaselliniCamardaBooth2022}.

Models~13 to 15 in Table~\ref{tab:models_4}, classified as functional time series models, employ nonparametric smoothing and functional data analysis techniques to produce accurate mortality forecasts while ensuring robustness against outliers. Specifically, the functional data model (FDM) applies functional principal component analysis (FPCA) to the entire mortality curve, allowing for a more comprehensive representation of mortality trends over time. The robust FDM extends the FDM by enhancing its robustness and mitigating the influence of outliers, resulting in more stable forecasts. Lastly, the product-ratio model of \cite{HBY13} adopts a mortality modeling approach based on the ratio of mortality functions across subgroups, such as female and male data, effectively capturing relative mortality patterns across populations. For the functional data approach, we consider the standard FDM for age-specific mortality, its robust variant to handle outliers, and coherent multi-population forecasting via the product–ratio method \citep{HyndmanBooth2008, HyndmanUllah2007, HBY13}.

The aforementioned models are widely used in mortality rate modeling and forecasting, each offering distinct advantages and limitations. The LC models are extensively used in actuarial and demographic applications due to their simplicity and interpretability. Functional time series models assume a smooth underlying stochastic process to model variations in age-specific mortality rates, thereby effectively capturing time-dependent mortality patterns.

Each family thus offers a different lens for modeling mortality changes -- some emphasize parametric assumptions (RH, CBD, and LC variants), whereas others adopt smoothing methods (FDM). Despite their methodological differences, these 15 models all rely on extrapolating historical patterns, making their forecasts somewhat correlated. Nevertheless, each model brings distinct strengths and weaknesses, underscoring the value of combining them through an ensemble approach. By starting with this diverse set, we can better assess how each contributes to forecast accuracy and, in turn, evaluate the advantages of the proposed ensemble methodology.

\subsection{Forecast Ensemble Approaches}\label{sec:2.2}

Ensemble modeling has become a cornerstone of modern forecasting applications, including mortality projections, due to its ability to combine multiple models and exploit their diverse predictive strengths. Rather than relying on a single "best" model, an ensemble can mitigate individual model weaknesses, reduce variance, and lower overall bias, ultimately improving forecast performance in highly complex settings \citep[see, e.g.,][]{bimonte2024mortality}.

A central challenge in ensemble construction is selecting appropriate weights for combining the constituent models. Methods such as simple model averaging (SMA) assign uniform weights to all models, ensuring simplicity but often failing to account for differences in model performance. Bayesian Model Averaging (BMA) is a more sophisticated approach that utilizes posterior probabilities to calculate weights, taking into account each model's fit to the data \citep[see, e.g.,][]{BenchimolEtAl2016, SH18}. In the mortality context, \cite{BenchimolEtAl2016} combine standard single-population models using both Akaike information criterion (AIC)-based weights and Bayesian model averaging, while \cite{DeMoriEtAl2024} analyze several model-averaging schemes in a two-population setting where the weights are based on each model's forecasting performance over a training period, which is conceptually close to the AIC-based weighting benchmark considered in this paper. 
While BMA accounts for model performance, it may struggle with overfitting when using highly complex models or when posterior distributions are sensitive to priors.

Although these techniques can provide solid baselines, their static nature can limit adaptability to shifting data patterns and varying forecast horizons, potentially leading to overfitting or failing to capture improvements from alternative models.

Recent developments in machine learning have led to more adaptive ensemble schemes, including stacked generalization \citep{wolpert1992stacked, KE22unified} and SHAP-based weighting \citep{lundberg2017unified}. Both belong to the ensemble family but differ in how they evaluate and combine individual model outputs.

Stacked generalization (often called stacking) trains a "meta-learner" on the predictions of the base models, adjusting their weights to minimize forecast error in a validation set. This approach is explicitly geared toward predictive accuracy, giving greater influence to models that reduce overall error.

SHAP-based weighting employs cooperative game theory to quantify each model's marginal contribution to the ensemble. Rather than directly optimizing for minimal error, it assigns "credit" based on how much each model improves the explainability or predictive power of the ensemble. This process can be repeated for different forecast horizons or data structures, enabling a more dynamic view of which models contribute most in specific contexts.

In mortality forecasting, these ensemble techniques are particularly valuable. Mortality data often exhibit intricate patterns over age, time, and cohort dimensions. Relying on a single model risks overlooking valuable information captured by others, while combining many models without carefully balancing their contributions may cause overfitting or computational inefficiency. By evaluating each model’s predictive performance and marginal contribution, ensemble methods can capture a broader range of trends, such as emerging cohort effects, age-specific improvements, and shifting overall mortality levels, while dampening noise and idiosyncratic errors.

We adopt a SHAP-based ensemble approach to refine mortality forecasts at multiple horizons. Unlike static weighting schemes that apply a single set of weights to all forecasts, our method recalculates each model's SHAP value at each horizon, dynamically adapting to temporal changes in model performance. Introducing a threshold for minimal SHAP values filters out models with little predictive utility, thus reducing computational complexity and enhancing stability.

The aim of this paper is twofold:
\begin{inparaenum}
\item[1)] We address whether an ensemble constructed using SHAP-derived optimal weights can outperform traditional model combination techniques in mortality forecasting. 
\item[2)] We explore how the bias-variance trade-off evolves when adopting SHAP weighting by comparing it to simpler weighting strategies, such as equal weighting.
\end{inparaenum}

By addressing these questions, the paper contributes to the literature in three distinct ways. 
\begin{inparaenum}[1)]
\item We propose an adaptive ensemble method that assigns dynamic weights to each mortality model based on its estimated contribution at different forecasting horizons. 
\item We empirically evaluate how SHAP-based ensemble weighting impacts forecast accuracy and variance reduction relative to traditional methods. 
\item Finally, we formalize the bias-variance decomposition for mortality model ensembles, illustrating how SHAP-based weighting improves predictive stability and robustness.
\end{inparaenum}

In Section~\ref{sec:3}, we present the theoretical underpinnings of this ensemble framework and illustrate how it balances accuracy, diversity, and interpretability, which are key attributes for reliable and robust mortality projections.

\section{Model Ensemble Criteria}\label{sec:3}

Out-of-sample evaluation techniques play a pivotal role in the design of effective ensembles. These methods ensure that model performance is assessed on data not used during training, allowing for a more reliable estimate of generalization error. Cross-validation, rolling-window evaluation, and block-based validation are techniques that provide frameworks for comparing models across different forecasting horizons and optimizing their contributions to the ensemble. Integrating out-of-sample performance metrics enables ensembles to address model uncertainty and effectively improve predictive accuracy across different forecasting horizons.

Our proposed SHAP ensemble approach distinguishes itself from the existing methods by dynamically calculating weights and assigning marginal contributions to each model, thereby optimizing both accuracy and diversity. Introducing a threshold parameter enables the exclusion of less relevant models, reducing overall variance without compromising accuracy.

We explore the key criteria for constructing effective ensembles, focusing on determining optimal weights, the role of thresholding in improving model quality, and the critical balance between bias and variance. This analysis demonstrates how SHAP offers a unique blend of accuracy and interpretability, delivering superior performance for both short-term and long-term forecasting horizons.

\subsection{Optimal Weighting Strategies}\label{sec:3.1}

Selecting optimal weights is crucial for enhancing the accuracy and robustness of ensemble models for mortality forecasting. The primary objective of weighting schemes is to achieve an appropriate balance among individual models, leveraging their complementary strengths while mitigating their weaknesses. A wide range of approaches to calculating weights have been explored in the literature, each with distinct advantages and limitations.

Empirical evidence indicates that methods incorporating optimal weighting schemes, such as stacked regression ensembles and SHAP-based ensembles, achieve superior long-term forecasting performance compared to simpler methods, such as simple model averaging (SMA) and AIC weights. The improved interpretability and flexibility of these advanced methods make them especially effective for addressing challenges in mortality data, ultimately supporting the production of precise and reliable forecasts for use in actuarial science and public health.

The SHAP values provide a game-theoretic framework for calculating weights by quantifying each model's contribution to the overall prediction at a highly detailed level, ensuring high interpretability. Moreover, incorporating a threshold parameter ensures that models with minimal contributions are excluded, thereby preventing the ensemble from being unduly influenced by underperforming methods. This approach balances computational efficiency and flexibility in adapting to varying data structures.

\subsection{Thresholding for SHAP-Based Ensemble Refinement}\label{sec:3.2}

The integration of a threshold mechanism within the ensemble weighting scheme is driven by the necessity to enhance the stability and efficiency of the forecasting model. While the SHAP framework effectively assigns weights based on each model's marginal contribution, it does not inherently distinguish between models that make meaningful contributions to the ensemble and those that make negligible or even detrimental contributions. As a result, including all models without filtering can lead to unnecessary complexity, increased variance, and reduced overall forecast reliability.

Using a threshold parameter enables the selective exclusion of models that make minimal contributions, thereby enhancing the ensemble's robustness. This filtering process mitigates the risk of overfitting by preventing models with high variance but low predictive power from excessively influencing the final forecast. Furthermore, implementing a threshold improves the ensemble's computational efficiency by reducing the number of active models, thereby facilitating scalability for high-dimensional forecasting problems.

Theoretically, applying a threshold can further optimize the trade-off between bias and variance. While excluding models with low contributions may slightly increase bias, it is expected to yield a net improvement in mean squared error (MSE) by significantly reducing variance. This is particularly relevant in mortality forecasting, where noisy models may introduce instability in long-term predictions. When applied dynamically across different forecasting horizons, the threshold mechanism ensures that the weighting scheme remains both interpretable and adaptable to varying data structures. This perspective is consistent with both the literature on optimal forecast combinations and practical insights into shrinkage and weighting in the presence of instability \citep{ElliottTimmermann2004, ElliottTimmermann2005, Timmermann2006ForecastCombinations, SmithWallis2009, GenreKennyMeylerTimmermann2013}.

Let the $h$-year-ahead mortality rate forecasts from $N$ mortality models $(M_1, \dots, M_N)$ be given by  $(\widehat{m}^{(1)}_{x,g,t+h}, \dots, \widehat{m}^{(N)}_{x,g,t+h})$ for age $x$ and gender $g$ at time $t+h$. The SHAP-weighted ensemble forecast is constructed as:
\begin{equation}\label{eq:1}
\widehat{m}^{(c)}_{\text{Shap}}(x,g,t+h) = \sum_{i=1}^{N} \widehat{\omega}_{i,\text{Shap}} \widehat{m}^{(i)}_{x,g,t+h},
\end{equation}
where $\widehat{\omega}_{i,\text{Shap}}$ represents the SHAP-derived weight for the $i$\textsuperscript{th} model and satisfies:
\begin{equation}
\widehat{\omega}_{i,\text{Shap}} = \frac{\widetilde{\phi}_i(\nu; g)}{\sum_{i=1}^{N} \widetilde{\phi}_i(\nu; g)}.
\end{equation}
where $\widetilde{\phi}_i(\nu; g)$ denotes the normalized mean SHAP value assigned to the $i$\textsuperscript{th} forecasting method, ensuring that models are weighted according to their estimated contribution to forecast accuracy.

To mitigate the influence of models with negligible contributions, we introduce a truncation mechanism that excludes models with mean SHAP values below a predefined threshold, $\alpha$. Let $S(\alpha)=\{\,i:\,\widetilde{\phi}_i(\nu;g)>\alpha\,\}$ be the selected set and
$N^\ast=|S(\alpha)|$. The truncated SHAP weights are
\begin{equation}
\widehat{\omega}_{i,\text{Shap}\,\alpha}
= \frac{\widetilde{\phi}_i(\nu;g)\,\mathds{1}\{i\in S(\alpha)\}}
       {\sum_{j\in S(\alpha)} \widetilde{\phi}_j(\nu;g)},
\end{equation}
where $\mathds{1}\{\cdot\}$ represents the binary indicator function. 
The SHAP-weighted ensemble with truncation is then given by:
\begin{equation}\label{eq:2}
\widehat{m}^{(c)}_{\text{Shap},\alpha}(x,g,t+h) = \sum_{i=1}^{N^*} \widehat{\omega}_{i,\text{Shap} \alpha} \widehat{m}^{(i)}_{x,g,t+h}.
\end{equation}
The threshold $\alpha$ determines the minimum mean SHAP value required for a model to contribute to the ensemble, with $N^*$ representing the number of models that satisfy  $\widetilde{\phi}_i(\nu; g) > \alpha$. 

The choice of threshold $\alpha$ is data-driven. For each country, sex, and horizon \(h\), we evaluate a small grid of candidate values (e.g., \(\{0.05, 0.10, 0.15, 0.20, 0.50\}\)) on the validation sample; for each $\alpha$ we compute the validation MSE and select the minimizer.

The truncation level $\alpha$ governs the bias–variance trade-off of the
ensemble. Small $\alpha$ (weak filtering) preserves diversity and typically stabilizes short-horizon variance; larger $\alpha$ (stronger filtering) prioritizes models with stronger conditional signal and can reduce bias at longer horizons, at the cost of less diversity. In this sense, thresholding is a shrinkage: a moderate cutoff stabilizes results, whereas an aggressive cutoff can raise bias and -- if it drops models that offset each other -- may even increase variance.

\subsection{The Bias-Variance Trade-off}\label{sec:3.3}

Ensemble methods can improve forecast accuracy by trading off bias and variance more effectively than individual models. Under squared-error loss, performance-weighted combinations approximate the regression benchmark when weights are reliably estimated, whereas in finite samples the variability of estimated weights can offset conditional gains—explaining why simple averages remain competitive at short horizons or under instability \citep{ElliottTimmermann2004, SmithWallis2009, GenreKennyMeylerTimmermann2013}. When relative model performance changes over time, adaptive or regime-aware weighting can be advantageous \citep{ElliottTimmermann2005}, while practical guidance and shrinkage ideas are surveyed in \citet{Timmermann2006ForecastCombinations}.

In this paper, we adopt SHAP-based regression ensembles, which quantify
marginal contributions to the combined forecast and employ a threshold to regulate diversity and stability (see Section~\ref{sec:3.2}). Small thresholds preserve diversity and tend to stabilize short-horizon variance; larger thresholds emphasize a strong conditional signal and may reduce long-horizon bias at the expense of diversity, in line with shrinkage principles discussed in the combination literature.

For completeness, formal definitions and the finite-sample bias–variance decomposition are given in~\ref{sec:app} (see Proposition~\ref{prop:cond} and Remark~\ref{rem:finite}).

\section{Implementation Details}\label{sec:4}

\subsection{Mortality data for OECD countries}\label{sec:4.1}

We consider age-specific mortality rates of 24 OECD countries obtained from the Human Mortality Database \citep{HMD24}: Austria, Belgium, Czech, Denmark, Estonia, Finland, France, Hungary, Iceland, Ireland, Italy, Japan, Latvia, Lithuania, Luxembourg, Netherlands, New Zealand, Norway, Poland, Spain, Sweden, Switzerland, UK, USA. Due to data availability, we select mortality observations for females and males of each country between 1960 and 2019. For ages, we have considered death rates for a single year of age, ranging from 0 to 100. Although the available data vary by country, we standardize the analysis to maintain consistency across this period. In particular, we split the total 60 years of data into a training set (1960-1999), a validation set (2000-2009), and a testing set (2010-2019).

\subsection{Ensemble mortality approach}\label{sec:4.2}

Combining forecasts from multiple models constitutes an alternative to relying on a single forecasting technique. The individual mortality models are combined using the approaches described in Section~\ref{sec:3}. These approaches empirically integrate predictions from various models, thus mitigating the uncertainty associated with model selection and enhancing point forecast accuracy \citep[see, e.g.,][]{bates1969combination}.

\subsubsection{SHAP weighting method}\label{sec:4.2.1}

After estimating each model, we aggregate the predictions to produce final combined forecasts. We use the SHAP approach to combine the multiple forecasts produced by the 15 methods listed in Tables~\ref{tab:models_1}-~\ref{tab:models_4}. Specifically, for each age $x$ and each forecasting horizon $h$, we compute SHAP values to measure the marginal contribution of each model, then combine them into a weighted forecast.

Let $x = x_1, \ldots, x_n$ denote all integer ages for a given gender $g$. For each age $x$:
\begin{enumerate}
\item[1)] Compute Base Forecasts: Fit each of the 15 mortality models to the training set and produce forecasts for horizons $h = 1, 2, \dots, 10$. Denote the forecast of the $i$\textsuperscript{th} method for age $x$, gender~$g$ in year $t$ by $\widehat{m}^{(i)}_{x,g,t}$. Collect all resulting forecasts (across $i = 1, \dots, 15$ and $h = 1, \dots, 10$) into a set of 15 features at each age $x$.
\item[2)] Calculate SHAP Values: For each method $i$, each age $x$, and horizon $h$, compute its SHAP value $\phi_{i}(\nu,x,g,h)$, which quantifies the ``importance'' or marginal contribution of model $i$'s forecast compared to the others.
\item[3)] Calculate mean SHAP Values: Let $\widetilde{\phi}_i(\nu; g)$ be the mean SHAP value associated with the $i\textsuperscript{th}$ forecasting method, normalized to the range $[0,1]$ automatically. 
\item[4)] \commHS{Construct the SHAP ensemble forecast by computing SHAP-derived weights that optimally quantify each model’s marginal contribution, in accordance with~\eqref{eq:1}.}
\item[5)] \commHS{Construct the SHAP-Combined Forecast with Truncation by introducing a mechanism that enhances computational efficiency according to~\eqref{eq:2}.}
\end{enumerate}

\subsubsection{Benchmark weighting methods}\label{sec:5.3.2}

To assess the accuracy of our newly proposed SHAP ensemble for mortality forecasting, we compare it against these weighting schemes:
\begin{itemize}
\item Simple model averaging: we apply equal weights of $\widehat{\text{average}}_i(\nu; g) = 1/15$ to individual forecasts obtained by each mortality model.
\item Akaike model averaging: we compute AIC for the out-of-sample forecasts produced by each method over the validation set. We then compute the {AIC-weights} as
\begin{equation*}
\widehat{\omega}_{i,\text{AIC}} = \frac{\widehat{\lvert\text{AIC}}_i(\nu; g)\rvert}{ \sum_{i=1}^{\bar{N}} \widehat{\lvert \text{AIC}}_i(\nu; g) \rvert } \mathds{1} \left\{ \widehat{ \text{AIC}}_i(\nu; g) < 0 \right\},
\end{equation*}
where $\widehat{\text{AIC}}_i(\nu; g)$ is the estimated AIC over the validation set for the $i\textsuperscript{th}$ method and $\bar{N}$ is the number of forecasting methods with negative estimate $\widehat{\text{AIC}}_i(\nu; g)$ values.
\end{itemize}

Using these benchmark weights, we combine the individual forecasts in analogy to~\eqref{eq:1}.

\section{Comparative Analysis of Point Forecast Accuracy}\label{sec:5}

After employing the validation set for weight selection, we use the test set to assess model performance through the MSE and mean absolute error (MAE). Specifically, to evaluate the point forecast accuracy, we compute the MSE and MAE for an ensemble of SHAP-weighted weights as:
\begin{align*}
\text{MSE}_{\text{Shap}} & = \frac{1}{101\times(11-h)} \sum_{\xi=h}^{10} \sum_{x=0}^{100} \left[  y_{x,g,2009+\xi} - \widehat{m}^{(c)}_{\text{Shap}}(x,g,2009+\xi)\right]^2,  \\
\text{MAE}_{\text{Shap}} & = \frac{1}{101\times(11-h)} \sum_{\xi=h}^{10} \sum_{x=0}^{100} \left| y_{x,g,2009+\xi} -  \widehat{m}^{(c)}_{\text{Shap}}(x,g,2009+\xi)\right|,
\end{align*}
where $y_{x,g,2009+\xi}$ is the log mortality rate for age $x$ and gender $g$ in year $2009+\xi$. 

The analysis focuses on both male and female populations and presents the percentage improvement in out-of-sample mortality forecast accuracy based on one-year-ahead to 10-year-ahead MSEs and MAEs. For brevity, in Table~\ref{table:shap_h1}, we report only the results for $h = 1$, $h = 6$, and $h = 10$-year-ahead forecasts of OECD countries’ mortality rates, measured by the average MSEs ($\times 100$). The results are shown for SHAP ensembles both without truncation and with a $50$\% truncation. To assess whether the SHAP ensembles, with and without truncation, perform competitively relative to the existing combined methods, we compare them to SMA and the AIC.
\begin{center}
\tabcolsep 0.145in
\renewcommand{\arraystretch}{0.78}
\begin{longtable}{@{}lcccccccc@{}}
\caption{One-step-ahead ($h=1$), six-step-ahead ($h=6$) and ten-step-ahead ($h=10$) MSE forecast accuracy of OECD mortality rates, for ensemble methods Average, AIC, SHAP and SHAP $\alpha=50\%$} \label{table:shap_h1} \\
\toprule 
& \multicolumn{2}{c}{SMA} & \multicolumn{2}{c}{AIC} & \multicolumn{2}{c}{SHAP} & \multicolumn{2}{c}{SHAP $\alpha=50\%$} \\
\cmidrule(lr){2-3} \cmidrule(lr){4-5} \cmidrule(lr){6-7} \cmidrule(lr){8-9}
Country & F & M & F & M & F & M & F & M \\
\midrule
\endfirsthead
\caption[]{(continued)} \\
\toprule 
& \multicolumn{2}{c}{SMA} & \multicolumn{2}{c}{AIC} & \multicolumn{2}{c}{SHAP} & \multicolumn{2}{c}{SHAP $\alpha=50\%$} \\
\cmidrule(lr){2-3} \cmidrule(lr){4-5} \cmidrule(lr){6-7} \cmidrule(lr){8-9}
Country & F & M & F & M & F & M & F & M \\
\midrule
\endhead
\midrule
\endfoot
$h=1$ & & & & & & & & \\
\cmidrule(r){1-1} 
Austria & 0.0049 & \textBF{0.0137} & \textBF{0.0044} & 0.0139 & 0.0048 & 0.0138 & 0.0052 & 0.0160 \\ 
Belgium & 0.0053 & \textBF{0.0223} & 0.0059 & 0.0233 & \textBF{0.0041} & 0.0254 & 0.0055 & 0.0484 \\ 
Czech & 0.0112 & 0.0573 & \textBF{0.0086} & \textBF{0.0517} & 0.0092 & 0.0647 & 0.0102 & 0.0755 \\ 
Denmark & 0.0046 & 0.0237 & 0.0037 & 0.0240 & \textBF{0.0033} & \textBF{0.0231} & 0.0045 & 0.0243 \\ 
Estonia & \textBF{0.0353} & 0.1364 & 0.0354 & \textBF{0.1347} & 0.0360 & 0.1446 & 0.0380 & 0.1759 \\ 
Finland & 0.0112 & \textBF{0.0532} & 0.0108 & 0.0535 & \textBF{0.0104} & 0.0551 & 0.0108 & 0.0544 \\ 
France & \textBF{0.0162} & 0.0159 & 0.0168 & \textBF{0.0159} & 0.0230 & 0.0217 & 0.0274 & 0.0260 \\ 
Hungary & 0.0092 & 0.0515 & \textBF{0.0086} & \textBF{0.0422} & 0.0100 & 0.0440 & 0.0114 & 0.0530 \\ 
Iceland & 0.0887 & \textBF{0.2617} & 0.0874 & 0.2640 & \textBF{0.0867} & 0.2628 & 0.0877 & 0.2727 \\ 
Ireland & \textBF{0.0111} & 0.0615 & 0.0115 & 0.0598 & 0.0131 & \textBF{0.0586} & 0.0149 & 0.0621 \\ 
Italy & 0.0035 & 0.0050 & 0.0032 & 0.0047 & \textBF{0.0029} & 0.0046 & 0.0035 & \textBF{0.0042} \\ 
Japan & 0.0012 & 0.0024 & 0.0012 & \textBF{0.0018} & \textBF{0.0010} & 0.0027 & 0.0016 & 0.0033 \\ 
Latvia & 0.0271 & 0.1028 & \textBF{0.0249} & 0.1036 & 0.0320 & 0.1032 & 0.0317 & \textBF{0.0920} \\ 
Lithuania & 0.0213 & 0.0808 & 0.0230 & 0.0765 & \textBF{0.0194} & 0.0706 & 0.0250 & \textBF{0.0625} \\ 
Luxemburg & 0.0962 & \textBF{0.3146} & 0.0979 & 0.3180 & 0.0978 & 0.3202 & \textBF{0.0874} & 0.3229 \\ 
Netherlands & 0.0038 & \textBF{0.0074} & 0.0037 & 0.0077 & \textBF{0.0032} & 0.0089 & 0.0039 & 0.0099 \\ 
New Zealand & 0.0182 & 0.0361 & 0.0162 & 0.0353 & 0.0102 & 0.0355 & \textBF{0.0085} & \textBF{0.0335} \\ 
Norway & 0.0073 & \textBF{0.0264} & 0.0086 & 0.0268 & \textBF{0.0065} & 0.0275 & \textBF{0.0065} & 0.0307 \\ 
Poland & \textBF{0.0033} & 0.0121 & 0.0051 & 0.0117 & 0.0057 & \textBF{0.0114} & 0.0074 & 0.0128 \\ 
Spain & 0.0033 & 0.0050 & \textBF{0.0027} & 0.0059 & 0.0030 & 0.0046 & 0.0036 & \textBF{0.0040} \\ 
Sweden & 0.0032 & \textBF{0.0104} & \textBF{0.0029} & 0.0108 & 0.0034 & 0.0106 & 0.0088 & 0.0144 \\ 
Switzerland & 0.0049 & \textBF{0.0169} & 0.0044 & 0.0171 & \textBF{0.0037} & 0.0171 & 0.0038 & 0.0293 \\ 
UK & 0.0031 & 0.0052 & 0.0026 & 0.0047 & \textBF{0.0023} & \textBF{0.0038} & 0.0025 & 0.0049 \\ 
USA & 0.0031 & 0.0029 & 0.0020 & 0.0026 & 0.0019 & \textBF{0.0016} & \textBF{0.0019} & 0.0019 \\ 
\midrule
Mean & 0.0166 & 0.0552 & \textBF{0.0163} & \textBF{0.0546} & 0.0164 & 0.0557 & 0.0172 & 0.0598 \\
\midrule
$h=6$ & & & & & & & & \\
\cmidrule(r){1-1}
  Austria & 0.0106 & 0.0368 & \textBF{0.0073} & \textBF{0.0160} & 0.0134 & 0.0222 & 0.0250 & 0.0169 \\ 
  Belgium & 0.0098 & 0.0364 & 0.0163 & \textBF{0.0274} & 0.0059 & 0.0464 & \textBF{0.0041} & 0.0783 \\ 
  Czech & 0.0199 & 0.0528 & \textBF{0.0071} & \textBF{0.0393} & 0.0180 & 0.0420 & 0.0201 & 0.0519 \\ 
  Denmark & 0.0046 & 0.0375 & \textBF{0.0032} & \textBF{0.0347} & 0.0089 & 0.0401 & 0.0303 & 0.0529 \\ 
  Estonia & 0.1200 & 0.1521 & 0.1208 & \textBF{0.1443} & \textBF{0.0779} & 0.1673 & 0.0863 & 0.2216 \\ 
  Finland & 0.0085 & 0.0647 & 0.0117 & \textBF{0.0559} & \textBF{0.0079} & 0.0631 & 0.0099 & 0.0569 \\ 
  France & 0.0218 & 0.0316 & \textBF{0.0086} & \textBF{0.0226} & 0.0325 & 0.0318 & 0.0323 & 0.0424 \\ 
  Hungary & 0.0360 & 0.0892 & \textBF{0.0143} & \textBF{0.0441} & 0.0229 & 0.0712 & 0.0293 & 0.1062 \\ 
  Iceland & 0.0721 & \textBF{0.2996} & 0.0662 & 0.3088 & \textBF{0.0638} & 0.3310 & 0.0696 & 0.3578 \\ 
  Ireland & 0.0288 & 0.1311 & \textBF{0.0177} & 0.0866 & 0.0242 & 0.0844 & 0.0414 & \textBF{0.0780} \\ 
  Italy & 0.0084 & 0.0225 & \textBF{0.0043} & \textBF{0.0071} & 0.0047 & 0.0184 & 0.0047 & 0.0148 \\ 
  Japan & \textBF{0.0022} & 0.0090 & 0.0061 & \textBF{0.0055} & 0.0025 & 0.0126 & 0.0059 & 0.0138 \\ 
  Latvia & 0.0232 & 0.0939 & \textBF{0.0196} & 0.0911 & 0.0236 & 0.0883 & 0.0407 & \textBF{0.0816} \\ 
  Lithuania & 0.1045 & \textBF{0.0838} & 0.1095 & 0.1220 & \textBF{0.0716} & 0.0959 & 0.0844 & 0.0998 \\ 
  Luxemburg & 0.1094 & \textBF{0.2107} & \textBF{0.0924} & 0.2127 & 0.0959 & 0.2252 & 0.0939 & 0.2713 \\ 
  Netherlands & 0.0117 & 0.0336 & 0.0086 & \textBF{0.0143} & \textBF{0.0079} & 0.0326 & 0.0126 & 0.0457 \\ 
  New Zealand & 0.0316 & 0.0476 & 0.0393 & \textBF{0.0369} & 0.0199 & 0.0411 & \textBF{0.0172} & 0.0377 \\ 
  Norway & 0.0165 & 0.0298 & 0.0168 & \textBF{0.0213} & 0.0091 & 0.0237 & \textBF{0.0073} & 0.0292 \\ 
  Poland & 0.0350 & 0.0956 & \textBF{0.0137} & \textBF{0.0184} & 0.0258 & 0.0373 & 0.0180 & 0.0227 \\ 
  Spain & 0.0099 & 0.0219 & 0.0046 & 0.0099 & 0.0058 & 0.0103 & \textBF{0.0039} & \textBF{0.0043} \\ 
  Sweden & \textBF{0.0043} & 0.0153 & 0.0103 & \textBF{0.0098} & 0.0083 & 0.0133 & 0.0207 & 0.0204 \\ 
  Switzerland & 0.0078 & 0.0420 & 0.0143 & \textBF{0.0145} & 0.0073 & 0.0286 & \textBF{0.0038} & 0.0492 \\ 
  UK & 0.0054 & 0.0215 & 0.0073 & \textBF{0.0070} & \textBF{0.0037} & 0.0141 & 0.0129 & 0.0192 \\ 
  USA & 0.0047 & 0.0204 & \textBF{0.0025} & \textBF{0.0024} & 0.0029 & 0.0043 & 0.0026 & 0.0031 \\
  \midrule
  Mean & 0.0295 & 0.0700 & 0.0259 & \textBF{0.0564} & \textBF{0.0235} & 0.0644 & 0.0282 & 0.0740 \\ 
  \midrule
  $h=10$ & & & & & & & & \\
\cmidrule(r){1-1}
Austria & 0.0424 & 0.0415 & 0.0184 & \textBF{0.0201} & 0.0133 & 0.0286 & \textBF{0.0109} & 0.0478 \\ 
  Belgium & 0.0268 & 0.0559 & 0.0244 & 0.0393 & 0.0087 & 0.0325 & \textBF{0.0026} & \textBF{0.0290} \\ 
  Czech & 0.0394 & 0.1398 & \textBF{0.0056} & \textBF{0.0570} & 0.0273 & 0.0968 & 0.0282 & 0.0932 \\ 
  Denmark & 0.0189 & 0.0559 & \textBF{0.0071} & \textBF{0.0520} & 0.0150 & 0.0693 & 0.0342 & 0.1355 \\ 
  Estonia & 0.0559 & \textBF{0.2116} & \textBF{0.0305} & 0.2166 & 0.0626 & 0.2683 & 0.0615 & 0.4564 \\ 
  Finland & 0.0241 & 0.1014 & 0.0179 & 0.0819 & \textBF{0.0133} & 0.1000 & 0.0153 & \textBF{0.0697} \\ 
  France & 0.0141 & 0.0302 & \textBF{0.0030} & \textBF{0.0166} & 0.0233 & 0.0282 & 0.0343 & 0.0437 \\ 
  Hungary & 0.0337 & 0.0460 & \textBF{0.0146} & \textBF{0.0264} & 0.0225 & 0.0355 & 0.0356 & 0.0656 \\ 
  Iceland & 0.0541 & 0.6384 & \textBF{0.0442} & \textBF{0.5906} & 0.0605 & 0.6538 & 0.0856 & 0.7037 \\ 
  Ireland & 0.0422 & 0.0983 & \textBF{0.0207} & 0.0987 & 0.0505 & 0.0821 & 0.1392 & \textBF{0.0801} \\ 
  Italy & 0.0198 & 0.0604 & 0.0119 & \textBF{0.0140} & \textBF{0.0068} & 0.0291 & 0.0085 & 0.0202 \\ 
  Japan & 0.0122 & 0.0357 & 0.0082 & \textBF{0.0048} & 0.0028 & 0.0147 & \textBF{0.0022} & 0.0068 \\ 
  Latvia & 0.0281 & 0.2381 & 0.0278 & 0.1573 & \textBF{0.0231} & 0.1756 & 0.0325 & \textBF{0.1208} \\ 
  Lithuania & 0.2487 & 0.2680 & 0.2108 & 0.2578 & 0.1455 & 0.2269 & \textBF{0.1140} & \textBF{0.1876} \\ 
  Luxemburg & 0.1436 & 0.2221 & 0.0858 & \textBF{0.0897} & 0.0995 & 0.1129 & \textBF{0.0803} & 0.1069 \\ 
  Netherlands & 0.0252 & 0.0380 & 0.0171 & \textBF{0.0212} & \textBF{0.0143} & 0.0724 & 0.0225 & 0.1638 \\ 
  New Zealand & 0.0546 & 0.1119 & 0.0584 & \textBF{0.0366} & \textBF{0.0316} & 0.0753 & 0.0382 & 0.0492 \\ 
  Norway & 0.0480 & 0.0498 & 0.0417 & \textBF{0.0218} & 0.0235 & 0.0374 & \textBF{0.0101} & 0.0773 \\ 
  Poland & 0.0436 & 0.0858 & \textBF{0.0231} & \textBF{0.0293} & 0.0277 & 0.0455 & 0.0312 & 0.0359 \\ 
  Spain & 0.0215 & 0.0403 & \textBF{0.0056} & 0.0078 & 0.0114 & 0.0061 & 0.0061 & \textBF{0.0056} \\ 
  Sweden & 0.0287 & 0.0445 & 0.0220 & \textBF{0.0232} & \textBF{0.0203} & 0.0385 & 0.0310 & 0.0557 \\ 
  Switzerland & 0.0285 & 0.0165 & 0.0314 & \textBF{0.0087} & 0.0155 & 0.0134 & \textBF{0.0064} & 0.0351 \\ 
  UK & 0.0207 & 0.0408 & \textBF{0.0076} & \textBF{0.0075} & 0.0096 & 0.0159 & 0.0186 & 0.0175 \\ 
  USA & 0.0158 & 0.0710 & 0.0072 & \textBF{0.0028} & \textBF{0.0058} & 0.0113 & 0.0086 & 0.0039 \\ 
  \midrule
  Mean & 0.0455 & 0.1142 & 0.0310 & \textBF{0.0784} & \textBF{0.0306} & 0.0946 & 0.0357 & 0.1088 \\ 
\end{longtable}
\end{center}

\vspace{-.4in}

\subsection{SHAP v.s. SMA and AIC benchmarks}\label{sec:5.1}

In examining the MSE and MAE results for males and females across 24 countries and forecast horizons $h = 1, \dots, 10$, we focused firstly on the SHAP-based ensemble and the SHAP-based ensemble truncated at $50$\%. In Table~\ref{table:shap_h1}, we compare them to the SMA and the AIC in terms of MSE. By way of illustration, Table~\ref{table:shap_h1} reports the MSE × 100 for one-, six-, and ten-year-ahead age-specific mortality forecasts obtained with four sets of combination weights. The bold numbers highlight the lowest MSEs among the evaluated ensemble methods. Below, we summarize the main patterns by forecast horizon, sex, and country.

At shorter horizons (\textbf{$h \approx 1\text{–}3$}), the SHAP-based ensemble improves on SMA in 16 of 24 female country cases (67\%) and 10 of 24 male cases (42\%). Gains over AIC are slightly weaker (58\% and 50\%, respectively). The truncated version is the best performer in only 10 female and seven male countries, illustrating that aggressive pruning yields limited benefits at the shortest horizon. Exceptions where SMA or AIC remain marginally superior include Denmark (males), Finland (females), and Austria (females).

Horizon length (\textbf{$h \approx 4\text{–}6$}) accentuates the advantage of SHAP weights. Relative to SMA, it achieves lower MSE in 18/24 (75\%) female and 16/24 (67\%) male cases. Against AIC, the pattern is asymmetric: SHAP prevails for females in 12 countries but for males in only 3. Across both sexes, the SHAP-based ensemble outperforms the truncated variant in roughly two-thirds of the countries; the latter remains useful in nations such as Spain and Norway, where its extra variance reduction offsets the small bias increase.

At the longest horizon (\textbf{$h \approx 7\text{–}10$}), the SHAP-based ensemble dominates SMA almost uniformly (females 20/24, males 20/24). The advantage over AIC broadens for females (13/24) but narrows sharply for males (4/24), reflecting the occasional long-range strength of AIC in male mortality (e.g., France, Denmark, Iceland). The truncated ensemble is now rarely the top option. Still, it attains the lowest MSE in a non-negligible subset (females in 16 countries and males in 14), notably in Belgium and Switzerland.

Regarding heterogeneity across countries, improvements are most consistent in Belgium, Italy, Norway, Spain, Sweden, and the United States, where SHAP (standard or truncated) outperforms both averaging schemes for every horizon and sex. Conversely, in countries with noisier or structurally volatile mortality (Estonia, Iceland, Lithuania), the relative ranking of the four combinations fluctuates markedly, underscoring the importance of adaptive weighting.

Similar to \cite{KE22unified}, our results show that the SHAP ensemble criteria effectively account for the impact of the forecast horizon when selecting the optimal mortality model. SHAP-based ensemble systematically enhances accuracy relative to equal weights, and the benefit grows with the forecast horizon. In contrast, AIC and SMA are not tied to a specific forecast horizon, which makes them less effective for choosing mortality models over longer forecast horizons. Moreover, although reducing low contribution models can be helpful, especially at very short horizons, the SHAP-based ensemble is the most robust configuration across horizons, sexes, and the majority of OECD countries examined.

\begin{figure}[!htb]
\centering
\subfloat[Norway Female]{\includegraphics[width=0.48\linewidth]{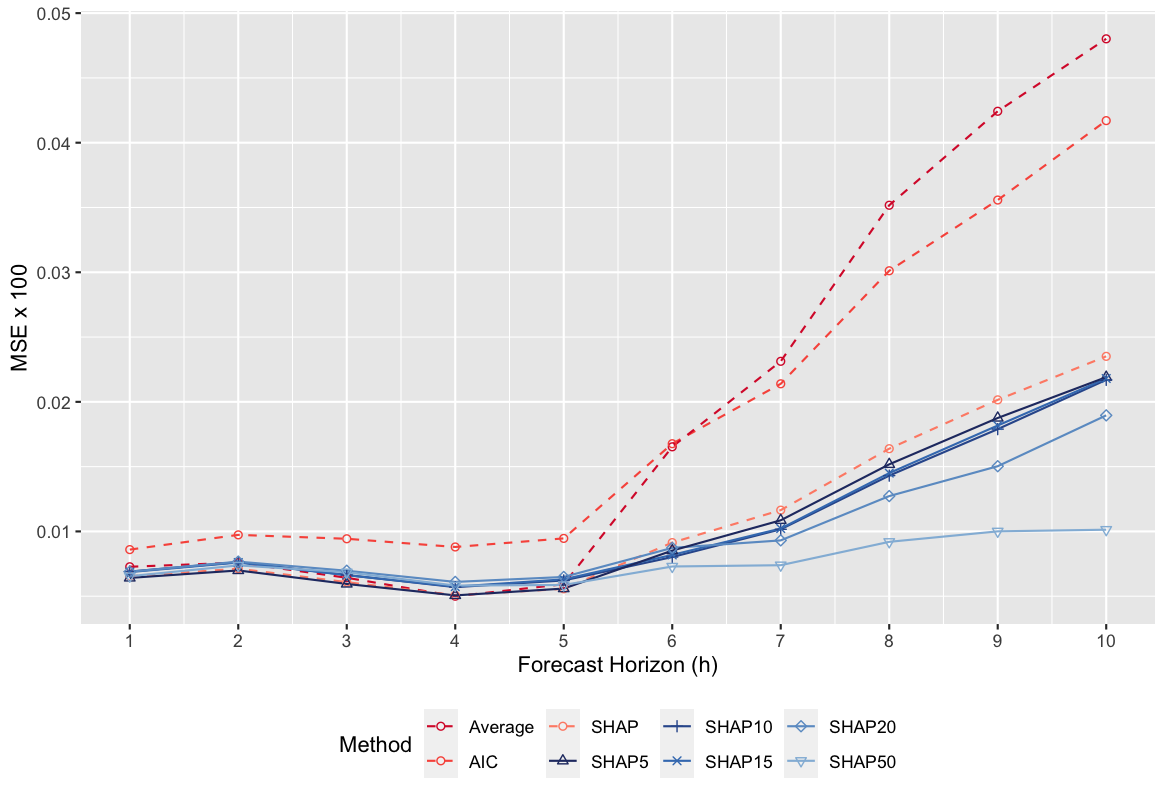}}
\hfill
\subfloat[Norway Male]{\includegraphics[width=0.48\linewidth]{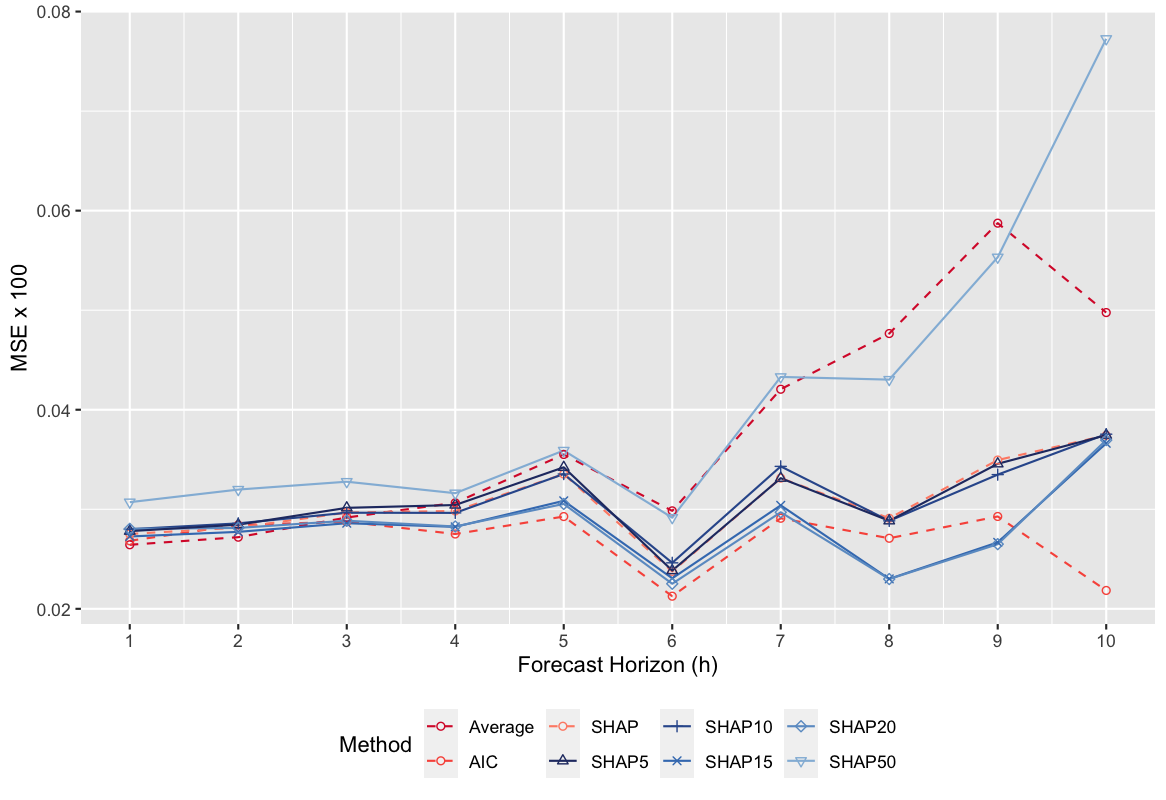}}
\par\smallskip
\subfloat[Spain Female]{\includegraphics[width=0.48\linewidth]{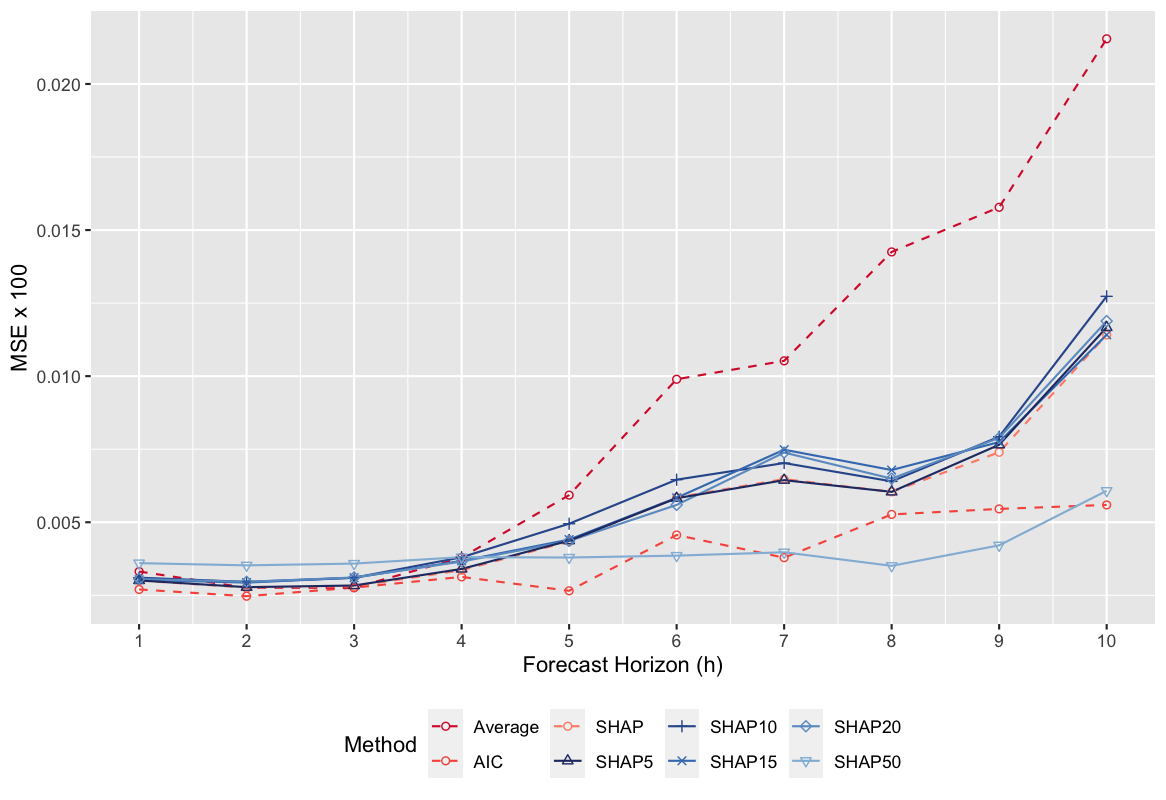}}
\hfill
\subfloat[Spain Male]{\includegraphics[width=0.48\linewidth]{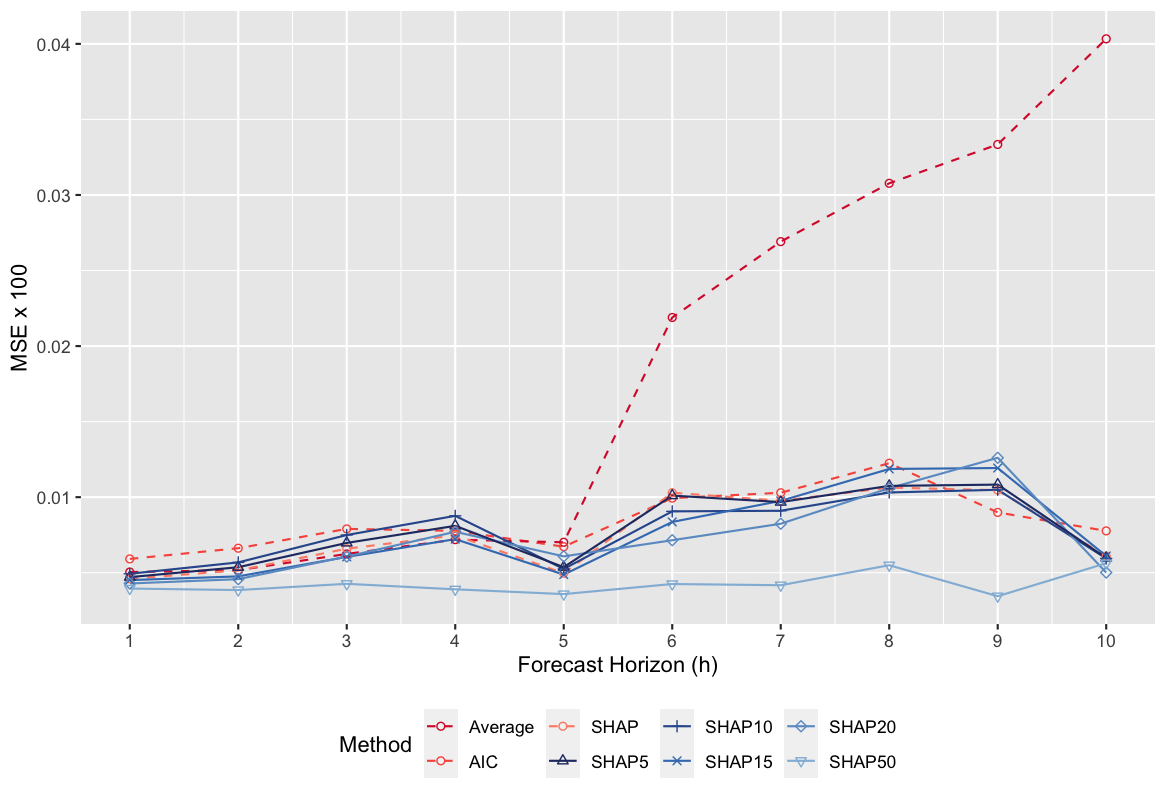}}
\par\smallskip
\subfloat[USA Female]{\includegraphics[width=0.48\linewidth]{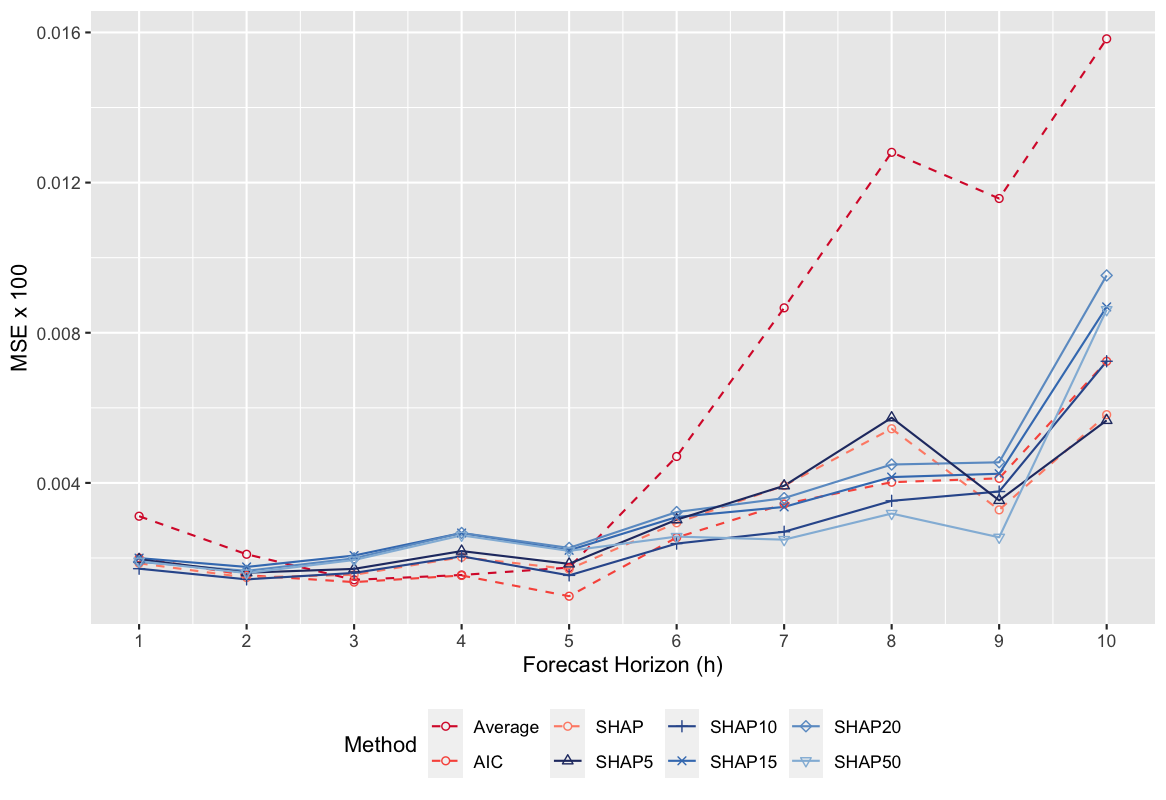}}
\hfill
\subfloat[USA Male]{\includegraphics[width=0.48\linewidth]{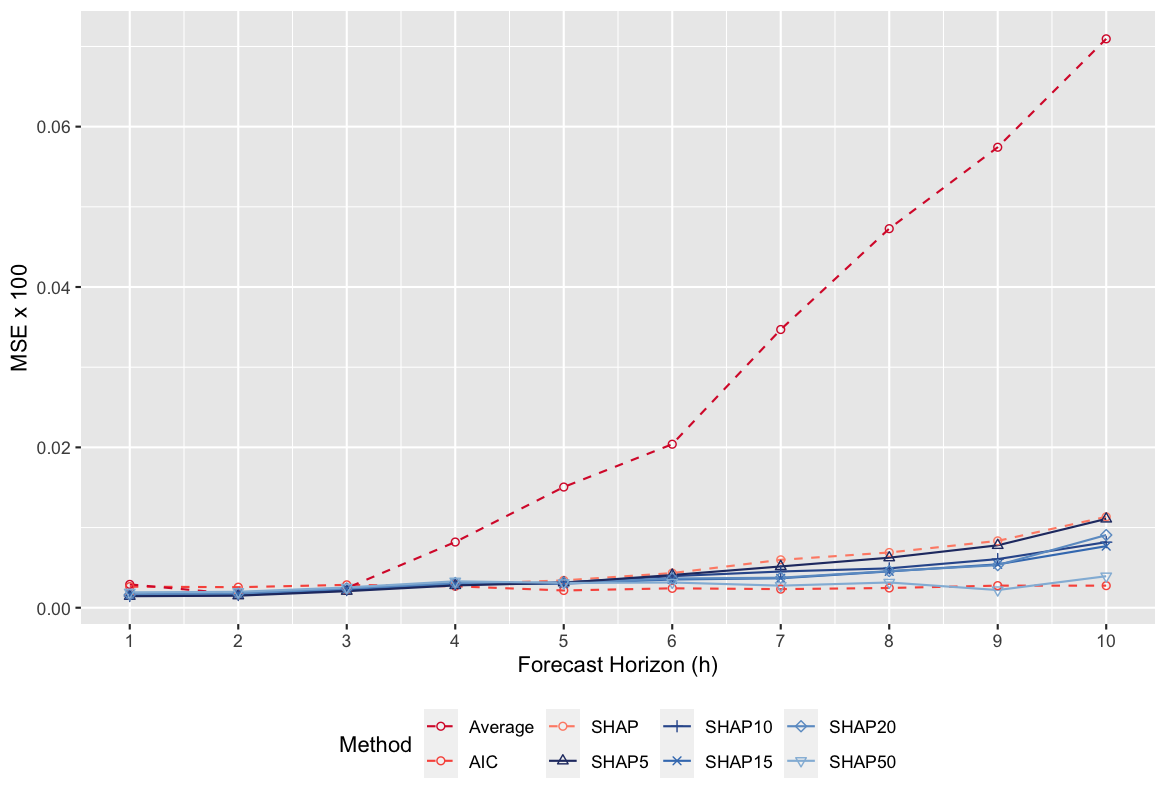}}
\caption{The SHAP ensembles on the female and male mortality data in Norway, Spain, and the USA for various horizons in terms of MSE.}
\label{fig:all_shap}
\end{figure}

Figure~\ref{fig:all_shap} illustrates all the SHAP ensemble forecasts for Norway, Spain, and the USA, confirming the consistency and robustness of SHAP-based strategies across different implementations. Similar results can also be observed for the other OECD countries, including for MAE, as shown in the Shiny App available at \url{https://github.com/YangANU/Ensemble_Mortality_Models.git}.

We also conduct the Diebold-Mariano test to assess the point forecast accuracy of SHAP, SMA, and AIC ensembles. For each integer age $x$ between 0 and 100, collate the out-of-sample point forecast errors $e_{\text{ensemble}} = y_{x,g,2009+\xi} - \widehat{m}^{(c)}_{\text{ensemble}}(x,g,2009+\xi)$ into a vector, where $h=1,\ldots10$, $\xi = h, \ldots,10$, and $\text{ensemble} \in \{ \text{SHAP}, \text{SMA}, \text{AIC}\}$. The 'dm.test' function from the 'forecast' package \citep{forecastR} in \Rlogo\ with a squared loss is then employed to assess the following hypotheses:
\begin{align*}
& H_{0} : \text{The two ensembles have the same forecast accuracy.} \\
& H_{A} : \text{The SHAP ensemble is more accurate than the benchmark.}
\end{align*}

\begin{figure}[!htb]
\centering
\includegraphics[width=0.85\linewidth]{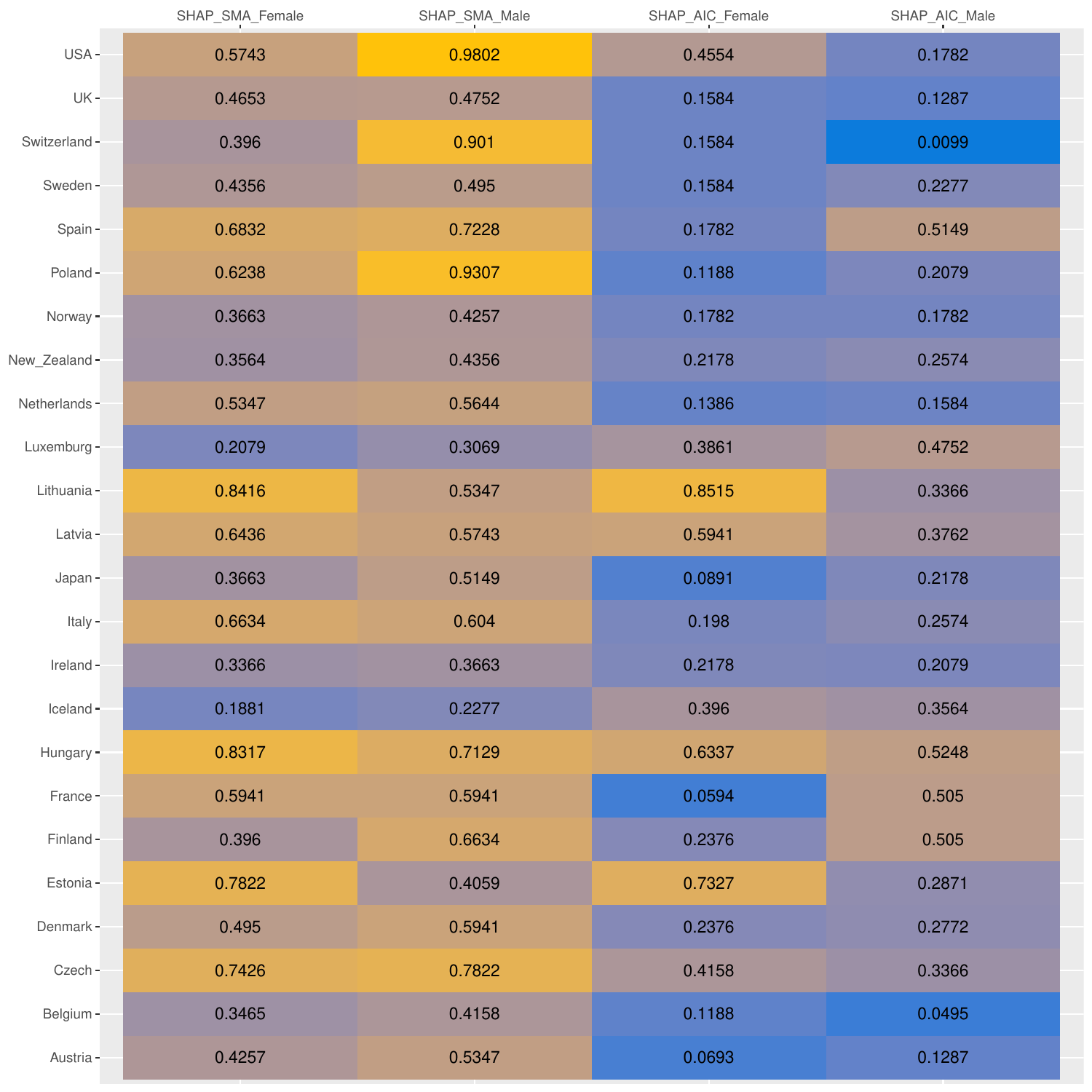}
\caption{\small Proportions of the SHAP ensemble outperforming the SMA and AIC methods across integer ages from 0 to 100, assessed using the Diebold-Mariano test for OECD countries. The values in blue are smaller than those in yellow.}\label{fig:dm_results}
\end{figure}

Adopting the commonly used 5\% significance level, we count the proportions of $p$-values smaller than $0.05$ out of the 101 integer ages for OECD countries and present the results in Figure~\ref{fig:dm_results}. 

The first two columns of the heatmap indicate the age ranges in which the SHAP ensemble outperforms the SMA method across OECD countries for both genders. Overall, the SHAP approach shows broad and robust gains: in most countries, it delivers more accurate forecasts for more than half of the ages for at least one gender. The last two columns report the share of ages for which the SHAP ensemble outperforms the AIC ensemble, highlighting that SHAP also provides improvements beyond AIC-based weighting. These gains are especially pronounced for the Baltic countries (Estonia, Lithuania, and Latvia).

\subsection{Age-stratified MSE}\label{sec:5.2}

\begin{figure}[!htb]
\centering
\includegraphics[width=0.8\linewidth]{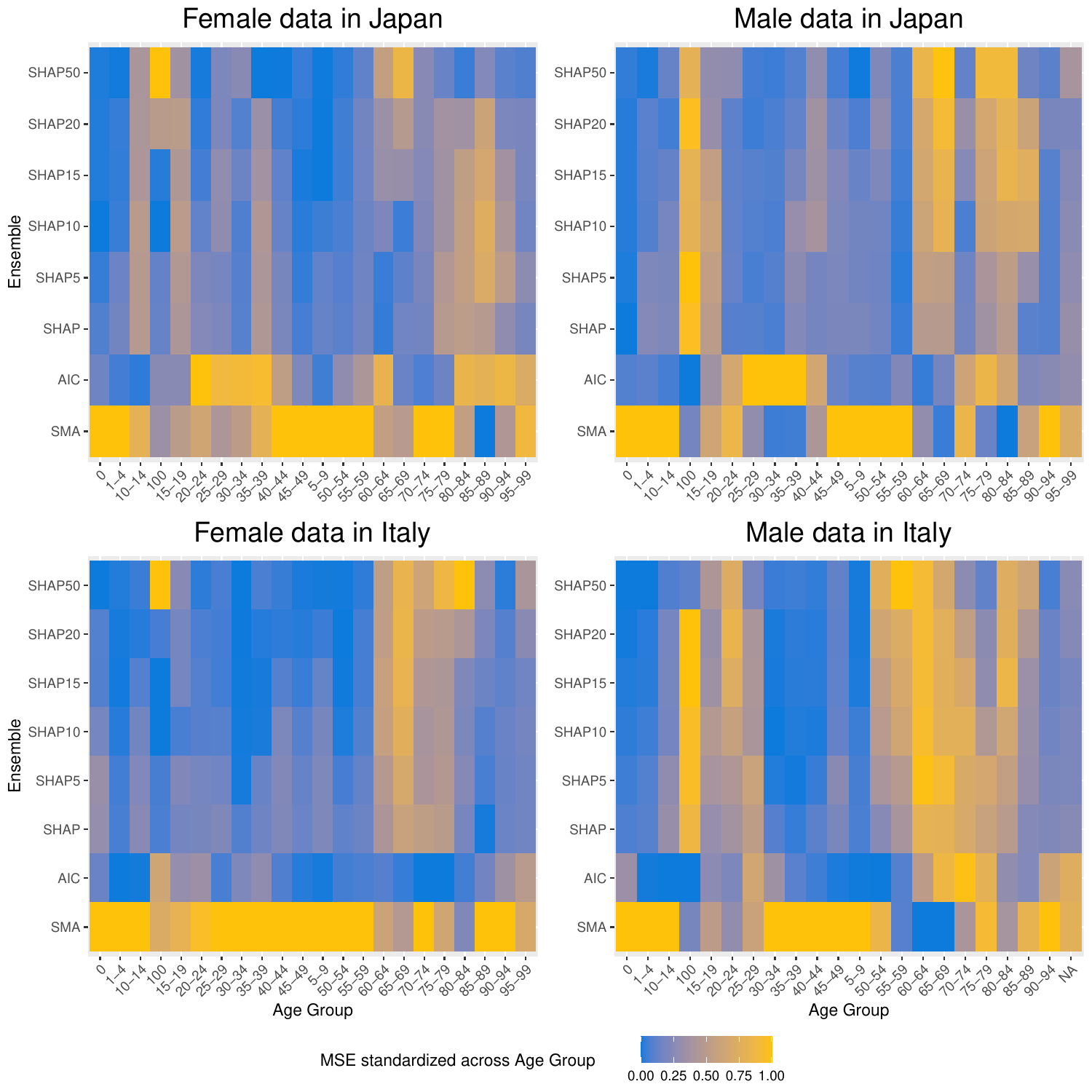}
\caption{The age-stratified MSE (standardized across age group) for female and male mortality rates in Japan and Italy.}
\label{fig:age_stratified_MSE}
\end{figure}

The heterogeneity in mortality patterns impacts the predictive performances of the considered ensembles. For example, the Plat model has been verified to work well for the Japanese mortality data \citep{SH18}, whereas the functional time series (i.e., FDM, robust FDM and product-ratio) models perform well for mortality rates in Italy \citep{bimonte2024mortality}. To highlight the ages where the SHAP ensemble performs strongly or poorly, we standardize the age-specific mortality forecasts from different ensembles for Japan and Italy and display the results as heatmaps in Figure~\ref{fig:age_stratified_MSE}. The age-stratified MSE heatmaps for other OECD countries can be accessed on the GitHub page.

The HMD age groups (i.e., 0, 1-4, 5-9, 10-14, $\ldots$, 91-94, 95-99, 100) \citep{HMD24} are used to prepare age-stratified MSE plots. For any age group, the largest MSE obtained by one of the eight ensemble methods will be standardized to 1 and shown in yellow. Similarly, the smallest MSE for each age is indicated by blue. Comparing the "yellow-blue" patterns in Figure~\ref{fig:age_stratified_MSE} indicates that SMA and AIC ensembles struggle to make accurate predictions for most ages less than 60 for the Japanese data. The SHAP ensembles with thresholding parameters over 10\% perform well on these ages. The SMA method provides accurate forecasts for elderly Japanese females in their 80s and for males between the ages of 75 and 90. In contrast, the SHAP-based and benchmark ensembles show slightly different patterns in Italian mortality rates. The SMA ensemble struggles on most ages for Italian females and in age intervals 0--20, 27--55, 80--83 for Italian males. The AIC ensemble yields superior forecasts for both genders in the 0--20 age range, for females between 60--80, and for males between 45--55. In Italy, the SHAP ensembles with $\alpha \ge 5\%$ perform better for both females and males aged 28--45.

\subsection{Sensitivity analysis}\label{sec:5.3}

\begin{figure}[!htb]
\centering
\begin{minipage}{0.48\textwidth}
\centering
\includegraphics[width=\linewidth]{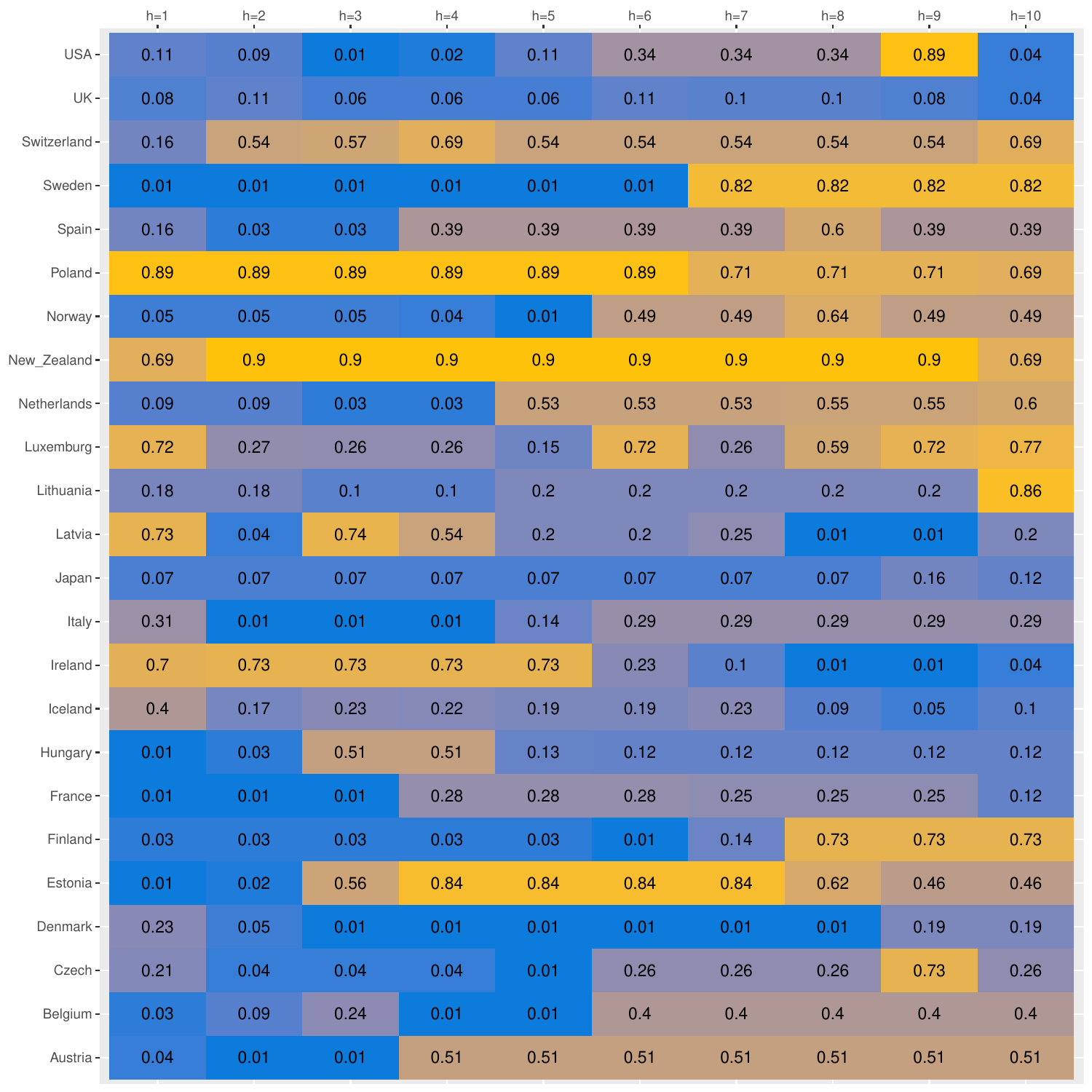}
\caption*{a - Female}
\end{minipage}
\quad
\begin{minipage}{0.48\textwidth}
\centering        
\includegraphics[width=\linewidth]{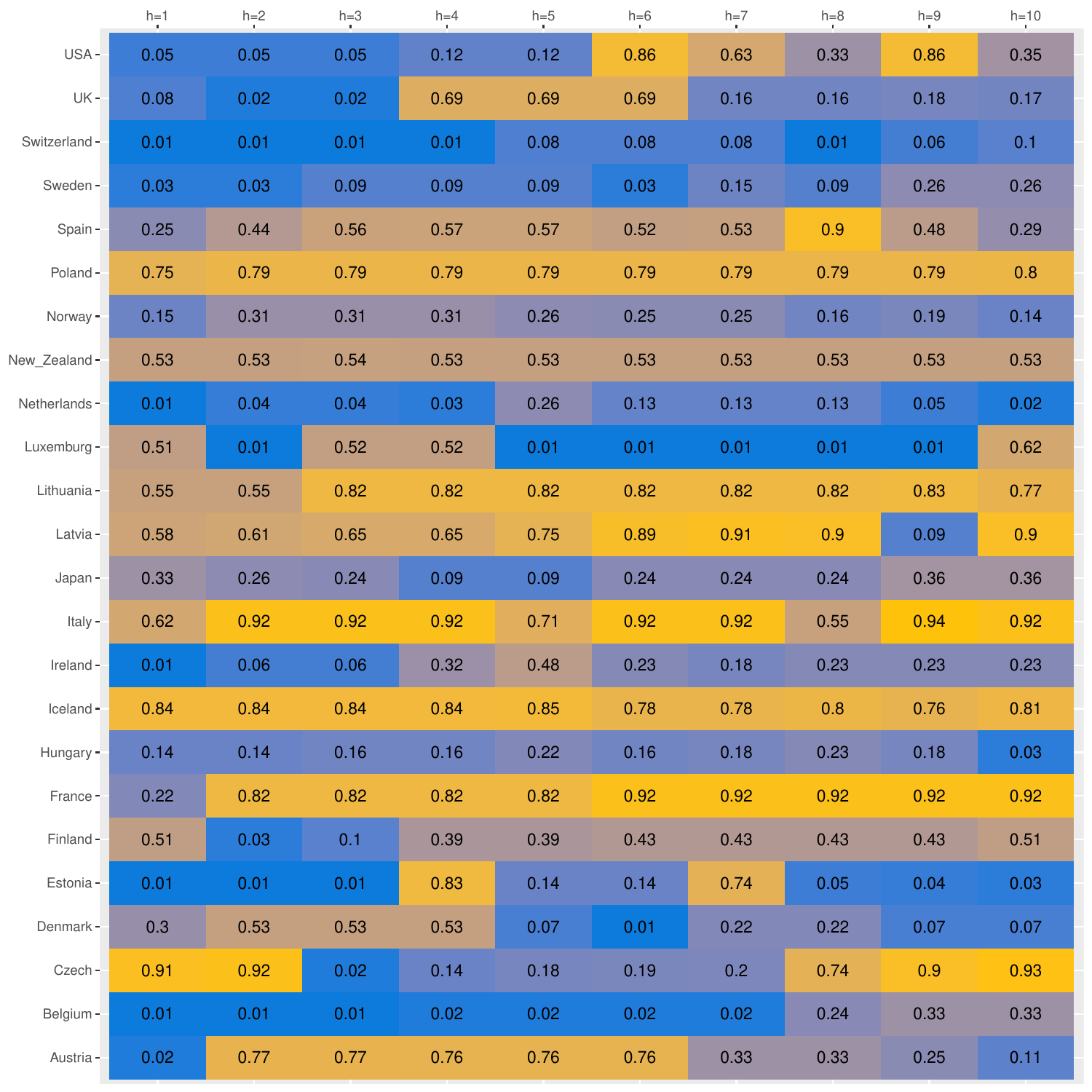}
\caption*{b - Male}
\end{minipage}    
\caption{The optimal truncation parameter $\alpha$ with the best point forecasting accuracy for each country selected by the grid search method for $h=1$ to $h=10$.}\label{heatmaps_alpha}
\end{figure}

To assess the sensitivity of truncated SHAP ensembles, we examined four values of the commonly used parameter $\alpha \in {0.05, 0.1, 0.15, 0.2, 0.5}$, as well as the parameter selected via grid search, across 24 OECD countries. The grid search method aims to identify the optimal $\alpha$ with the best point forecasting accuracy for each country. Specifically, we repeat the computation for $\alpha \in \{0.01,0.02,\ldots,0.99\}$ and evaluate the obtained SHAP ensembles. Figure~\ref{heatmaps_alpha} shows the optimal $\alpha$ estimates for $h=1$ to $h=10$.

It can be seen that larger truncation parameters are generally required for longer forecasting horizons. This is because forecasts for the distant future incorporate more uncertainties. Moreover, it is common to observe a higher $\alpha$ for males than for females because male mortality rates are more volatile with greater variances.

We repeat the computation separately for males and females and present the MSE for $h=1$ forecasts in Table~\ref{table:shap_sens_2}. Comprehensive results across all horizons ($h = 1$ to $10$), including MAE metrics, are available in the online Shiny App on GitHub. We notice that at shorter horizons (\textbf{$h \approx 1\text{–}3$}), the threshold $\alpha=0.05$ for SHAP values delivers the lowest MSE in 16 of 24 female series and 14 of 24 male series. In other words, keeping almost all models while cutting the weakest contributions yields the largest short-term gains, outperforming not only the ensemble based on SHAP without threshold, but also SMA, AIC, and SHAP with a $ 50\%$ threshold.

\begin{center}
\tabcolsep 0.25in
\renewcommand{\arraystretch}{0.83}
\begin{longtable}{@{}lcccccccc@{}}
\caption{\small One-step-ahead ($h=1$) forecast accuracy of OECD mortality rates, for SHAP ensemble with $\alpha$ selected by the grid search method and $\alpha \in \{0.05, 0.1, 0.15, 0.2, 0.5\}$.} \label{table:shap_sens_2} \\
\toprule 
& \multicolumn{2}{c}{\textbf{Grid Search}} & \multicolumn{2}{c}{\textbf{SHAP $\alpha=5\%$}} & \multicolumn{2}{c}{\textbf{SHAP $\alpha=10\%$}}  \\
\cmidrule(lr){2-3} \cmidrule(lr){4-5} \cmidrule(lr){6-7}
Country & F & M & F & M & F & M  \\
  Austria & \textBF{0.0047} & \textBF{0.0138} & 0.0049 & 0.0139 & 0.0051 & 0.0142  \\ 
  Belgium & \textBF{0.0041} & \textBF{0.0254} & 0.0042 & 0.0256 & 0.0044 & 0.0265 \\ 
  Czech & \textBF{0.0090} & \textBF{0.0511} & 0.0091 & 0.0650 & 0.0092 & 0.0651 \\ 
  Denmark & \textBF{0.0032} & \textBF{0.0228} & 0.0033 & 0.0231 & 0.0033 & 0.0233 \\ 
  Estonia & \textBF{0.0360} & \textBF{0.1446} & 0.0361 & 0.1451 & 0.0362 & 0.1478 \\ 
  Finland & \textBF{0.0104} & \textBF{0.0542} & 0.0104 & 0.0551 & 0.0105 & 0.0553 \\ 
  France & \textBF{0.0230} & \textBF{0.0207} & 0.0236 & 0.0224 & 0.0236 & 0.0226 \\ 
  Hungary & \textBF{0.0100} & \textBF{0.0416} & 0.0100 & 0.0442 & 0.0104 & 0.0440 \\ 
  Iceland & \textBF{0.0863} & \textBF{0.2523} & 0.0867 & 0.2628 & 0.0868 & 0.2631 \\ 
  Ireland & \textBF{0.0119} & \textBF{0.0586} & 0.0132 & 0.0588 & 0.0134 & 0.0592 \\ 
  Italy & \textBF{0.0027} & \textBF{0.0037} & 0.0029 & 0.0046 & 0.0029 & 0.0046 \\ 
  Japan & \textBF{0.0009} & \textBF{0.0022} & 0.0010 & 0.0028 & 0.0009 & 0.0026 \\ 
  Latvia & \textBF{0.0238} & \textBF{0.0906} & 0.0322 & 0.1036 & 0.0340 & 0.1014 \\ 
  Lithuania & \textBF{0.0187} & \textBF{0.0612} & 0.0193 & 0.0704 & 0.0191 & 0.0702 \\ 
  Luxemburg & \textBF{0.0855} & \textBF{0.3199} & 0.0975 & 0.3235 & 0.1012 & 0.3246 \\ 
  Netherlands & \textBF{0.0032} & \textBF{0.0089} & 0.0032 & 0.0090 & 0.0032 & 0.0091 \\ 
  New Zealand & \textBF{0.0073} & \textBF{0.0330} & 0.0099 & 0.0355 & 0.0095 & 0.0355 \\ 
  Norway & \textBF{0.0064} & \textBF{0.0273} & 0.0064 & 0.0278 & 0.0069 & 0.0280 \\ 
  Poland & \textBF{0.0045} & \textBF{0.0099} & 0.0058 & 0.0114 & 0.0060 & 0.0118 \\ 
  Spain & \textBF{0.0030} & \textBF{0.0039} & 0.0030 & 0.0047 & 0.0031 & 0.0049 \\ 
  Sweden & \textBF{0.0034} & \textBF{0.0105} & 0.0034 & 0.0105 & 0.0034 & 0.0108 \\ 
  Switzerland & \textBF{0.0034} & \textBF{0.0171} & 0.0037 & 0.0171 & 0.0037 & 0.0173 \\ 
  UK & \textBF{0.0021} & \textBF{0.0037} & 0.0022 & 0.0040 & 0.0022 & 0.0037 \\ 
  USA &\textBF{ 0.0017} & \textBF{0.0014} & 0.0020 & 0.0014 & 0.0017 & 0.0015 \\
  \cline{2-7} 
  Mean & \textBF{0.0152} & \textBF{0.0533} & 0.0164 & 0.0559 & 0.0167 & 0.0561 \\ 
  \midrule
  & \multicolumn{2}{c}{\textbf{SHAP $\alpha=15\%$}} & \multicolumn{2}{c}{\textbf{SHAP $\alpha=20\%$}} & \multicolumn{2}{c}{\textbf{SHAP $\alpha=50\%$}}  \\
\cmidrule(lr){2-3} \cmidrule(lr){4-5} \cmidrule(lr){6-7}
Country & F & M & F & M & F & M  \\
\midrule
  Austria & 0.0053 & 0.0143 & 0.0053 & 0.0147 & 0.0052 & 0.0160 \\ 
  Belgium & 0.0045 & 0.0280 & 0.0048 & 0.0285 & 0.0055 & 0.0484 \\ 
  Czech & 0.0090 & 0.0649 & 0.0090 & 0.0650 & 0.0102 & 0.0755 \\ 
  Denmark & 0.0033 & 0.0240 & 0.0033 & 0.0239 & 0.0045 & 0.0243 \\ 
  Estonia & 0.0371 & 0.1464 & 0.0374 & 0.1475 & 0.0380 & 0.1759 \\ 
  Finland & 0.0105 & 0.0559 & 0.0105 & 0.0561 & 0.0108 & 0.0544 \\ 
  France  & 0.0244 & 0.0217 & 0.0248 & 0.0207 & 0.0274 & 0.0260 \\ 
  Hungary & 0.0107 & 0.0418 & 0.0122 & 0.0428 & 0.0114 & 0.0530 \\ 
  Iceland & 0.0866 & 0.2638 & 0.0867 & 0.2649 & 0.0877 & 0.2727 \\ 
  Ireland & 0.0128 & 0.0598 & 0.0128 & 0.0618 & 0.0149 & 0.0621 \\ 
  Italy & 0.0029 & 0.0046 & 0.0030 & 0.0047 & 0.0035 & 0.0042 \\ 
  Japan & 0.0009 & 0.0026 & 0.0012 & 0.0027 & 0.0016 & 0.0033 \\ 
  Latvia & 0.0360 & 0.1023 & 0.0363 & 0.1021 & 0.0317 & 0.0920 \\ 
  Lithuania & 0.0196 & 0.0695 & 0.0195 & 0.0690 & 0.0250 & 0.0625 \\ 
  Luxemburg & 0.1005 & 0.3256 & 0.1009 & 0.3250 & 0.0874 & 0.3229 \\ 
  Netherlands & 0.0032 & 0.0097 & 0.0036 & 0.0100 & 0.0039 & 0.0099 \\ 
  New Zealand & 0.0088 & 0.0355 & 0.0088 & 0.0351 & 0.0085 & 0.0335 \\ 
  Norway & 0.0069 & 0.0273 & 0.0069 & 0.0280 & 0.0065 & 0.0307 \\ 
  Poland & 0.0061 & 0.0127 & 0.0060 & 0.0124 & 0.0074 & 0.0128 \\ 
  Spain & 0.0030 & 0.0045 & 0.0031 & 0.0043 & 0.0036 & 0.0040 \\ 
  Sweden & 0.0035 & 0.0111 & 0.0035 & 0.0110 & 0.0088 & 0.0144 \\ 
  Switzerland & 0.0036 & 0.0174 & 0.0036 & 0.0176 & 0.0038 & 0.0293 \\ 
  UK & 0.0022 & 0.0038 & 0.0022 & 0.0041 & 0.0025 & 0.0049 \\ 
  USA & 0.0020 & 0.0016 & 0.0019 & 0.0016 & 0.0019 & 0.0019 \\ 
  \cline{2-7}
  Mean & 0.0168 & 0.0562 & 0.0170 & 0.0564 & 0.0172 & 0.0598 \\
 \bottomrule
\end{longtable}
\end{center}

\vspace{-.4in}

However, at longer horizons ($h \geq 4$), the SHAP ensemble truncated at $50$\% emerges as the best performer across many countries, outperforming or closely matching the newly tested truncated SHAP methods and averaging-based approaches. While some country and sex specific nuances exist, for example, in Italy and Switzerland, the $\alpha=0.05$ truncated SHAP ensemble stands out for females at shorter horizons. In contrast, the $\alpha=0.10$ variant performs better for males in Denmark and Spain; no single truncation level dominates across all scenarios. Noticeably, the optimal parameter selected by grid search provides marginal improvements over the commonly used $\alpha \in {0.05, 0.1, 0.15, 0.2}$ at the cost of slow computation speed.
 
These findings underscore the added flexibility and robust predictive accuracy gained by refining the SHAP ensemble method with varying truncation thresholds while highlighting the competitiveness of the $50$\% truncated SHAP in longer-term forecasts.

\subsection{Challenges in Mortality Model Selection}\label{sec:5.4}

\begin{figure}[!htb]
\centering
\begin{minipage}{0.48\textwidth}
\centering
\includegraphics[width=\linewidth]{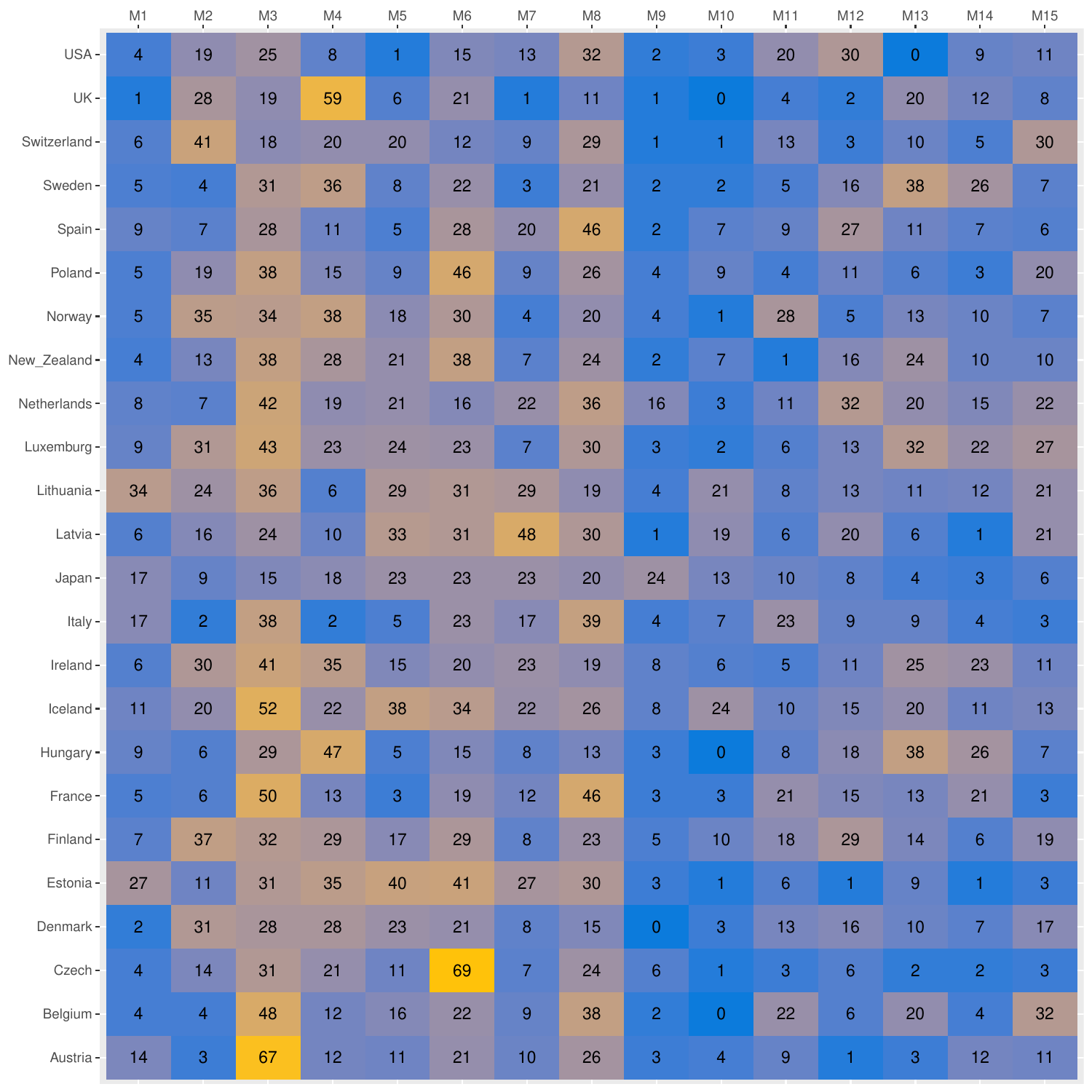}
\caption*{a - Female}
\end{minipage}
\quad
\begin{minipage}{0.48\textwidth}
\centering        
\includegraphics[width=\linewidth]{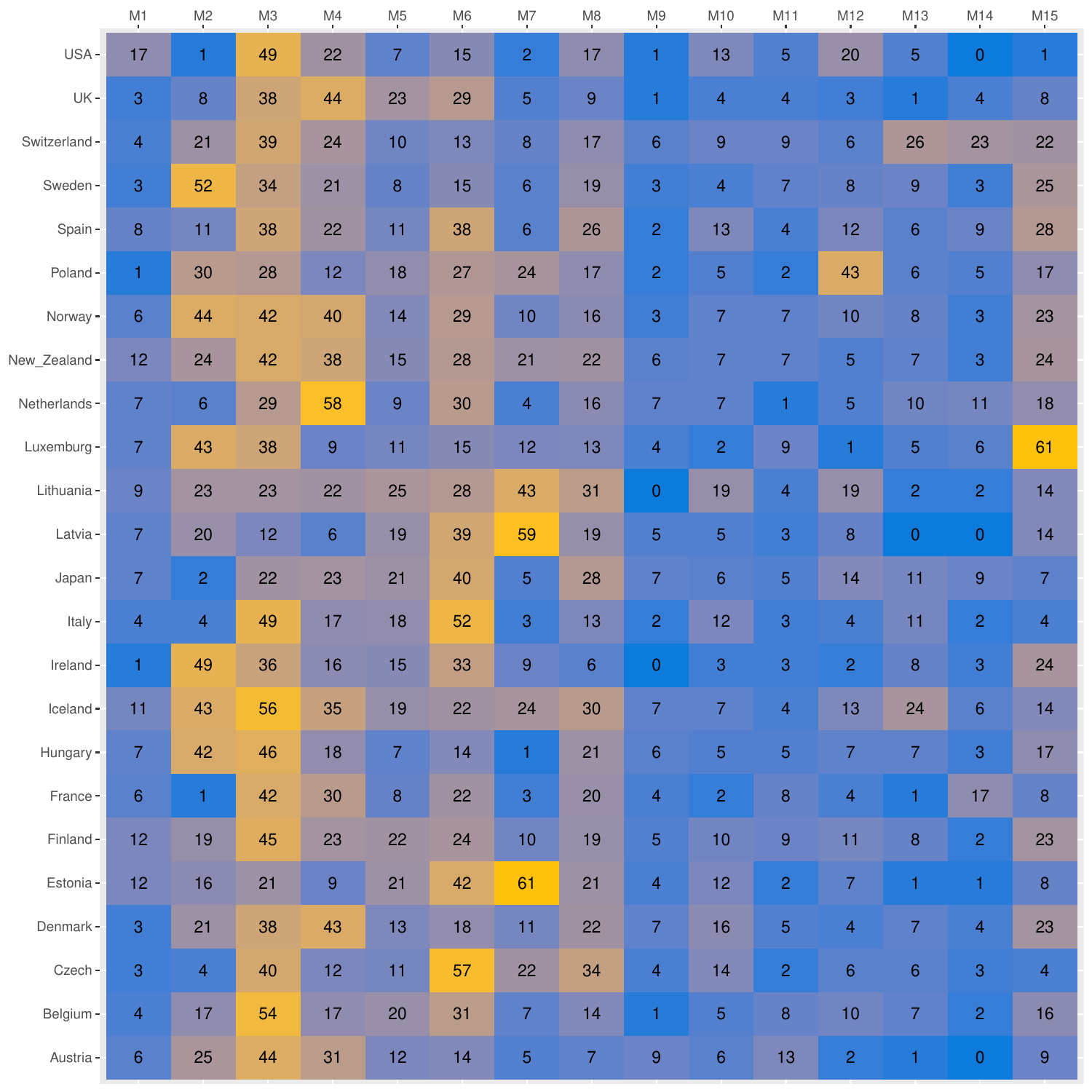}
\caption*{b - Male}
\end{minipage}    
\caption{\small Base forecasting methods after excluding those with the bottom 50\% SHAP values. The "M1" to "M15" represent "lc", "rh", "apc", "cbd", "m6", "m7", "m8", "plat", "lca\_dt", "lca\_dxt", "lca\_e0", "lca\_none", "fdm", "robust\_fdm", and "pr" base forecasting methods. The numbers in the two heatmaps correspond to the number of times each method was selected across 101 iterations; the maximum in each cell is 101, corresponding to the integer age 0-100.}\label{heatmaps}
\end{figure}

Figure~\ref{heatmaps} shows the heatmaps of the base forecast methods after excluding those with SHAP values below $50$\% for male and female populations. For simplicity, the numbers in the two heatmaps correspond to the number of times each method was selected across 101 iterations, i.e., one iteration of the SHAP50 algorithm for each age between 0 and 100. If a base forecasting method performs well, this frequency will be close to 101. Otherwise, it will approach zero.

For the female population, the analysis reveals notable geographical patterns in mortality modeling, with countries often showing concentrated strengths in particular methods while displaying weaker or more dispersed results in others. In the Nordic countries (Sweden, Norway, Finland, Iceland, Denmark), for example, there is a consistent strength in M3. However, Denmark and Norway complement it with Renshaw–Haberman  M2, while Iceland excels in M5 and M6. All five countries share low M12 fits. 

The Baltic States (Estonia, Latvia, and Lithuania) favour cohort-sensitive variants of the CBD family. Estonia distributes its high-frequency selections across M3, M4, M5, M6, and M8; Latvia focuses almost entirely on the block M5–M8; while Lithuania combines the Lee–Carter baseline M1 with M3 and M6. All three countries assign virtually no weight to the functional model M13.

Central Europe (Poland, Czech Republic, Hungary) shows a clear preference for the Plat extension M6 next to M3; Hungary alone enriches the mix with the functional model M13. 

In Western Europe (France, Belgium, the Netherlands, Luxembourg), models M3 and M8 dominate the selections. Luxembourg and Belgium further attribute substantial weight to the functional M13 and the product-ratio M15, whereas France assigns almost no weight to M15. 

Among the Anglophone countries (USA, UK, Ireland, and New Zealand), APC M3 is universally prominent. Yet, each nation favours a distinct secondary model: the USA boosts the score‐adjusted Lee–Carter M12, the UK leans on CBD M4, Ireland pairs M2 with M4, and New Zealand shows a marked preference for the Plat variant M6. 

Japan stands out with consistently low fits across most methods, suggesting unique characteristics or poor applicability of these approaches.

For males, M3 is the core model, yet two clear patterns surface. The basic CBD model M4 matches, or even surpasses, M3 across much of northern Europe (Austria, Denmark, Iceland, the Netherlands, Norway, Sweden, UK). Also, the strongly cohort-driven variants M6 and M7 dominate the Baltic trio (Estonia, Latvia, Lithuania) and often appear alongside M3 in Belgium, Czechia, Italy, and Spain. Luxembourg is the only country where the M15 product ratio model joins M2 and M3 in the top group, while Poland pairs M2 with the M12 adjusted for life expectancy.

\commHS{It is worth stressing that models M4--M6 belong to the Cairns--Blake--Dowd family, which was originally intended for old-age mortality. In this paper, we apply these CBD-type models to the full 0--100 age range to maintain both the comparability across base models and the homogeneity of the ensemble construction. As a consequence, the interpretation of these specifications is most meaningful at adult and older ages. In contrast, at younger ages, their potential misspecification is mitigated by the presence of alternative model families in the ensemble. The favourable performance of M4--M6 in some countries should therefore be read as evidence that, despite their old-age origin, these models can still provide useful information for the combined forecasts, especially at older ages, while other models (such as Lee--Carter variants and the functional models M13 and M14) tend to dominate the contribution at younger ages.}

The observed differences across country-specific models highlight the challenge of identifying consistent preferences in mortality modeling. Similarly, it is difficult to identify a single mortality prediction method that reliably performs across all contexts. In addition, the forecast performance of selected models varies depending on the gender used to fit the data, with no single model consistently performing best for males and females. These cross-country and gender-based variations underscore the challenge of pinpointing a single, universally optimal approach to mortality forecasting. Indeed, forecasting performance differs by country and gender, and no single model consistently outperforms the others under all conditions. This complexity underscores the decision to employ an ensemble of models, which captures the diverse strengths of each method and enhances overall predictive accuracy.

\subsection{Ensembles vs. single models}\label{sec:5.5}

A comparison of SHAP ensembles and individual models based on MSE shows that ensembles generally outperform single models across several countries for both males and females. It is noteworthy that this advantage is evident in New Zealand, Norway, Spain, and the USA, where SHAP-based forecasts consistently enhance predictive accuracy across all horizons and sexes.

In contrast, in several other countries, the advantage of SHAP ensembles appears to be gender-specific. For instance, the performance of SHAP ensembles is notably superior for females in Belgium, Italy, Japan (a particularly remarkable result given the peculiarities of the Japanese data), Lithuania, and Switzerland. In comparison, countries such as the Czech Republic, Finland, Ireland, Latvia, and the United Kingdom exhibit a male-specific advantage, in which ensemble methods substantially enhance forecasts for male mortality.

\begin{figure}[!htb]
\centering
\subfloat[Female data in Japan]{
\includegraphics[width=0.95\linewidth]{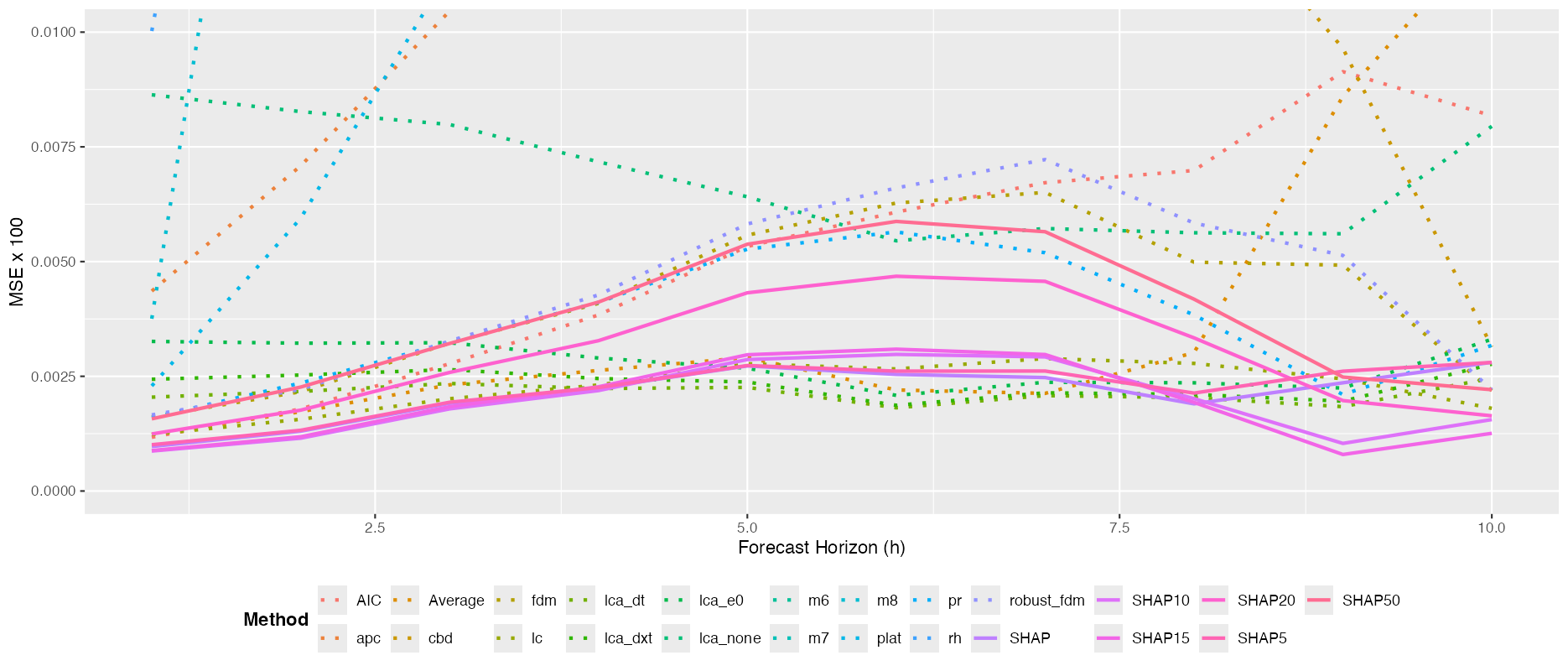}
}
\hfill
\subfloat[Male data in Ireland]{
\includegraphics[width=0.95\linewidth]{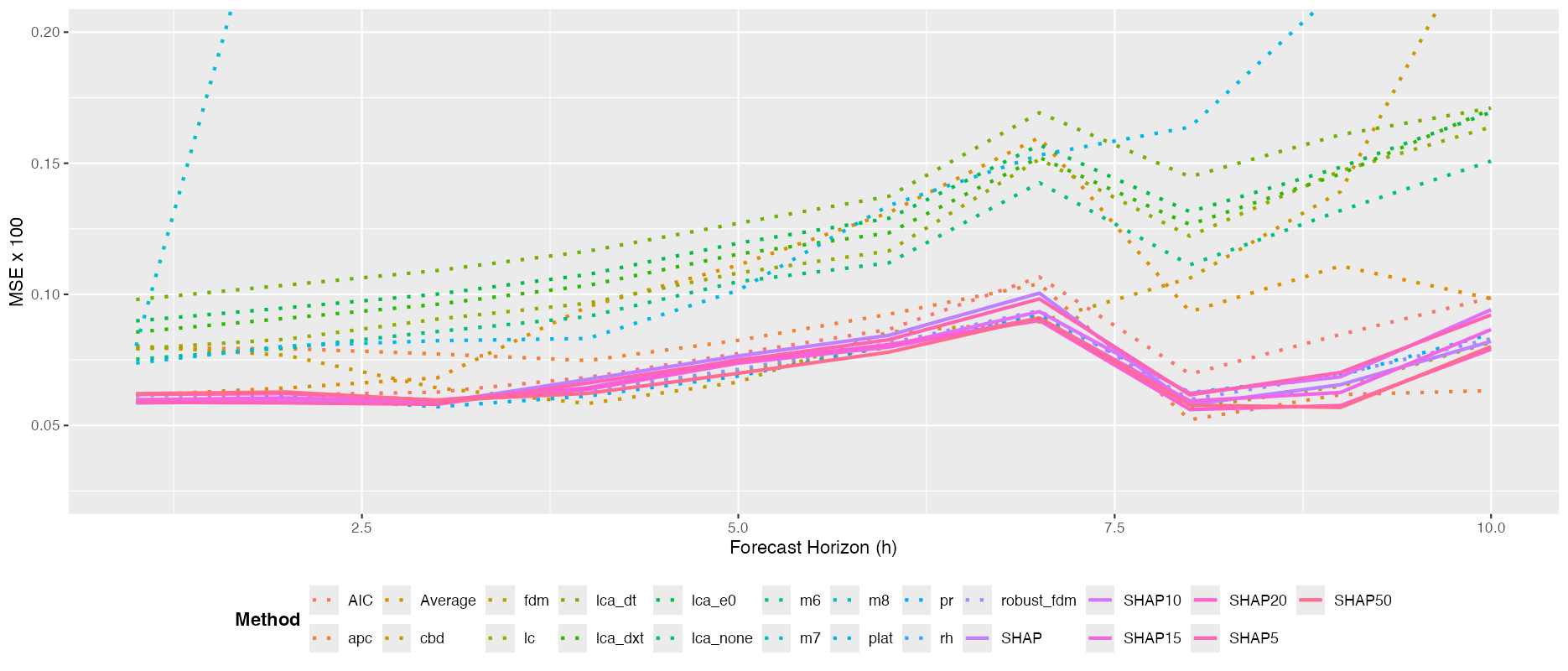}
 }
\caption{\small SHAP ensembles vs. single-model forecasts for Japan (female) and Ireland (male).}\label{fig:SHAP_vs_single}
\end{figure}

The findings emphasise that while SHAP ensembles offer a significant and often substantial performance improvement, no individual model, or combination of models, is universally optimal across all countries and both sexes. These results align with the insights of \cite{KE22unified}, who highlight the inherent challenge in identifying a dominant mortality forecasting approach that consistently performs well across diverse demographic and geographic contexts. The observed heterogeneity underscores the importance of model ensembles and the prudence of ensemble-based strategies, especially in the presence of country- and gender-specific dynamics. Figure~\ref{fig:SHAP_vs_single} illustrates this comparison by showing the SHAP-based ensemble forecasts versus individual model forecasts for two representative cases: Japan (female) and Ireland (male), where ensemble methods notably outperform the benchmarks. The results for each model, analyzed using MSE, are available in a Shiny App on GitHub.

\section{Ensemble Prediction Interval} \label{sec:6}

It is crucial to quantify the uncertainty associated with mortality forecasting methods in practical applications. We follow \cite{SH18} to compute out-of-sample prediction intervals for the testing data from 2010 to 2019 using the expanding window approach. Similar to the point forecast analysis, this method produces 10 one-step-ahead intervals, 9 two-step-ahead intervals, $\ldots,$ and 1 ten-step-ahead interval. We then construct the $h$-step-ahead ensemble prediction intervals following the methods introduced in the next section.

\subsection{Ensemble interval construction}

We consider four prediction interval combination methods proposed by \cite{grushka2020combining} and \cite{li2025ensemble}, simple averaging (SA), interior trimming (IT), AIC, and MSE, along with an ensemble interval construction approach based on SHAP-derived weights.

Let $(L_i,U_i)$ denote the central $100(1-\alpha)\%$ prediction interval for a particular mortality forecasting model with $i = 1,2,\ldots,N$. Let $(L_C,U_C)$ be the ensemble $100(1-\alpha)\%$ prediction interval constructed with method $C \in \{SA,\ IT, \ AIC, \ MSE, \ SHAP\}$ on the $N$ prediction intervals computed from the individual forecasting models.

\subsubsection{Simple averaging}\label{sec:6.1.1}

The simple averaging method computes the lower and upper endpoints across all models under consideration. Essentially, an identical weight of $1/N$ is applied to each mortality forecasting model. This ensemble interval construction method has proven useful in many applications despite its simple form.
\begin{equation*}
L_{SA} = \frac{1}{N}\sum_{i=1}^N L_i, \qquad
U_{SA} = \frac{1}{N}\sum_{i=1}^N U_i.
\end{equation*}

\subsubsection{Interior trimming}\label{sec:6.1.2}

The interior trimming method excludes the $d$ largest lower endpoints and the $d$ smallest upper endpoints, and then calculates the simple average of the remaining prediction intervals. Let the subscript $(j)$ denote the $j$th order statistic, the ensemble interval is computed as
\begin{equation*}
L_{IT} = \frac{1}{N-d}\sum_{j=1}^{N-d} L_{(j)}, \qquad
U_{IT} = \frac{1}{N-d}\sum_{j=d+1}^N U_{(j)}.
\end{equation*}
In effect, this method assigns a weight of 0 to the trimmed models and an equal weight to each remaining prediction interval. We follow \cite{li2025ensemble} and \cite{grushka2020combining} in excluding 20\% models, that is, $d = 15\times0.2 = 3$ models in this analysis.

\subsubsection{Weighted averaging based on AIC}\label{sec:6.1.3}

This method computes the weighted average of the lower and upper endpoints over models with negative AIC estimates, that is, $\widehat{\text{AIC}}_i(\nu; g) < 0$, over the validation set. Since lower AIC values indicate better model fitting, the absolute signs ensure that higher weights are assigned to these models.
\begin{equation*}
    L_{AIC} = \sum_{i=1}^N w_{i,\text{AIC}} L_i, \qquad
    U_{AIC} = \sum_{i=1}^N w_{i,\text{AIC}} U_i,
\end{equation*}
where $N^{\star}$ is the number of forecasting methods with negative estimate $\widehat{\text{AIC}}_i(\nu; g)$ values, and
\begin{equation*}
    w_{i,\text{AIC}}  = \frac{\widehat{\lvert\text{AIC}}_i(\nu; g)\rvert}{ \sum_{i=1}^{N^{\star}} \widehat{\lvert\text{AIC}}_i(\nu; g)\rvert } \mathds{1} \left\{ \widehat{\text{AIC}}_i(\nu; g) < 0 \right\}.
\end{equation*}

\subsubsection{Weighted averaging based on MSE}\label{sec:6.1.4}

Instead of using the fitting performance to do averaging as above, the MSE of forecasts over the validation set can also be used to select the weights $w_i$ as follows.
\begin{align*}
L_{MSE} &= \sum_{i=1}^N w_{i,\text{MSE}} L_i, \qquad w_{i,\text{MSE}} = \frac{\exp(-\text{MSE}_i)}{ \sum_{i=1}^{N} \exp(-\text{MSE}_i)}, \\
U_{MSE} &= \sum_{i=1}^N w_{i,\text{MSE}} U_i.
\end{align*}
The exponential operation ensures that weights decay modestly as MSE increases, leading to higher weights for models with good point forecasting accuracy.

\subsubsection{Weighted averaging based on SHAP}\label{sec:6.1.5}

The SHAP-based weighting approach can also be applied to construct ensemble prediction intervals. The SHAP weighting scheme selects weights based on the normalized mean SHAP values across the validation set and assigns larger weights to models with lower bias and variance. For a particular age, we compute the SHAP weights and the prediction interval given by
\begin{equation*}
L_{SHAP} = \sum_{i=1}^N w_{i,\text{SHAP}} L_i, \qquad
U_{SHAP} = \sum_{i=1}^N w_{i,\text{SHAP}} U_i,
\end{equation*}
where $N^{*}$ represents the number of models with SHAP values above the threshold $\alpha$, and
\begin{equation*}
w_{i,\text{SHAP}} = \frac{\widetilde{\phi}_i(\nu; g)}{\sum_{i=1}^{N^{*}} \widetilde{\phi}_i(\nu; g)}\mathds{1}\left\{\widetilde{\phi}_i(\nu; g) > \alpha \right\}.
\end{equation*}

\subsection{Evaluation of interval forecasts}\label{sec:6.2}

We evaluate the interval forecast accuracy via the interval score of \cite{gneiting2007strictly} \citep[see also][]{gneiting2014probabilistic}. The pointwise prediction interval score at the logarithm scale is defined as
\begin{align*}
  S_{\theta}\big[L_{C}(x,g,t+h), U_{C}(x,g,t+h), & y_{x,g,t+h}\big] =\  \left[U_{C}(x,g,t+h)-L_{C}(x,g,t+h)\right] \\
& + \frac{2}{\theta}\left[L_{C}(x,g,t+h) - \ln y_{x,g,t+h}\right]\mathds{1}\left\{\ln y_{x,g,t+h}<L_{C}(x,g,t+h)\right\}\\
& + \frac{2}{\theta}\left[\ln y_{x,g,t+h} - U_{C}(x,g,t+h)\right]\mathds{1}\left\{\ln y_{x,g,t+h} > U_{C}(x,g,t+h)\right\}.
\end{align*}
The interval score rewards a narrow prediction interval if and only if the holdout observation lies within the prediction interval. The optimal interval score is achieved when the true observation lies between $L_{C}(x,g,t+h)$ and $U_{C}(x,g,t+h)$, and the pointwise distance between $L_{C}(x,g,t+h)$ and $U_{C}(x,g,t+h)$ is minimal. The parameter $\theta$ denotes the level of significance and is customarily selected as $\theta = 0.2$. For various forecasting horizons $h$ and different time points in the mortality curve, the mean interval score is defined by
\begin{align*}
\overline{S}_{\theta}(h) = \frac{1}{101\times(11-h)} \sum_{\xi=h}^{11} \sum_{x=0}^{101}  S_{\theta}\left[L_{C}(x,g,t+\xi), U_{C}(x,g,t+\xi), y_{x,g,t+\xi}\right].
\end{align*}

In Table~\ref{table:interval_score_h1}, we present the mean interval scores for one-step-ahead forecasts, using the simple average, interior trimming, AIC, MSE, and the SHAP-based weighting scheme. The moderate truncation level of $\alpha = 10\%$ is selected as an example of thresholding the SHAP ensemble. Evaluation results for two- to ten-step-ahead interval forecasts are available in the Shiny App on GitHub. 
\begin{center}
\tabcolsep 0.25in
\renewcommand{\arraystretch}{0.83}
\begin{longtable}{@{}lcccccccc@{}}
\caption{\small One-step-ahead ($h=1$) mean interval score of ensemble interval forecasts for OECD mortality countries. This table reports results associated with simple average, SHAP without truncation, SHAP with $\alpha = 10\%$, interior trimming, AIC-based, and MSE-based ensembles.} \label{table:interval_score_h1} \\
\toprule 
& \multicolumn{2}{c}{\textBF{Simple Average}} & \multicolumn{2}{c}{\textBF{SHAP}} & \multicolumn{2}{c}{\textBF{SHAP $\alpha=10\%$}}  \\
\cmidrule(lr){2-3} \cmidrule(lr){4-5} \cmidrule(lr){6-7}
Country & F & M & F & M & F & M  \\
\midrule
\endfirsthead
\midrule
\endfoot
\endlastfoot
Austria     & 1.6864 & 1.5280 & 1.6700 & 1.5548 & 1.7342 & 1.7093 \\ 
Belgium     & 1.6788 & 1.4757 & 1.6132 & 1.1635 & 1.6001 & 1.0953 \\ 
Czech       & 1.3585 & 1.4590 & 1.3957 & 1.5281 & 1.5512 & 1.5946 \\ 
Denmark     & 2.1237 & 1.7903 & 1.7273 & 1.5503 & 1.7465 & 1.5655 \\ 
Estonia     & 1.8764 & 1.7977 & 1.7750 & 2.0704 & 1.8542 & 2.1110 \\ 
Finland     & 1.6775 & 1.5688 & 1.6757 & 1.6257 & 1.7041 & 1.6412 \\ 
France      & 1.2599 & 1.2731 & 0.8499 & 0.8595 & 0.8113 & 0.8075 \\ 
Hungary     & 2.0375 & 2.3011 & 1.4091 & 1.5939 & 1.4158 & 1.6545 \\ 
Iceland     & 1.9836 & 2.2088 & 2.2782 & 2.3339 & 2.3502 & 2.3652 \\ 
Ireland     & 2.0303 & 1.7180 & 1.8859 & 1.8936 & 1.8862 & 1.7325 \\ 
Italy       & 1.3691 & 1.1925 & 1.8284 & 1.1631 & 1.9453 & 1.1845 \\ 
Japan       & 1.0573 & 1.0303 & 1.1275 & 0.9539 & 1.1946 & 1.0273 \\ 
Latvia      & 2.7536 & 1.9807 & 2.8336 & 2.2892 & 2.8907 & 2.3254 \\ 
Lithuania   & 1.7458 & 2.3068 & 1.6090 & 2.2937 & 1.6261 & 2.2891 \\ 
Luxemburg   & 1.7800 & 2.0453 & 1.9208 & 2.0565 & 2.0115 & 2.0951 \\ 
Netherlands & 1.6632 & 1.4559 & 1.3372 & 1.3839 & 1.2902 & 1.6189 \\ 
New Zealand & 1.6344 & 1.5191 & 1.8735 & 1.7124 & 1.9834 & 1.7703 \\ 
Norway      & 1.7496 & 1.4510 & 1.5948 & 1.5384 & 1.6026 & 1.5721 \\ 
Poland      & 1.5216 & 1.5150 & 1.6181 & 1.3784 & 1.8231 & 1.3634 \\ 
Spain       & 1.3867 & 7.8200 & 1.3840 & 5.2977 & 1.3233 & 3.0897 \\ 
Sweden      & 1.4622 & 1.5585 & 1.3448 & 1.3330 & 1.3817 & 1.3352 \\ 
Switzerland & 1.9896 & 1.9386 & 1.8089 & 1.7400 & 1.8106 & 1.8183 \\ 
UK          & 1.3987 & 1.4396 & 1.6277 & 1.6355 & 2.0662 & 1.7464 \\ 
USA         & 1.4975 & 1.4247 & 1.0614 & 0.8744 & 0.9944 & 0.6934 \\
\cmidrule{2-7}
Mean        & 1.6967 & 1.9083 & 1.6354 & 1.7427 & 1.6589 & 1.7101 \\ 
\midrule
 & \multicolumn{2}{c}{\textBF{Interior Trimming}} & \multicolumn{2}{c}{\textBF{AIC}} & \multicolumn{2}{c}{\textBF{MSE}} \\
\cmidrule(lr){2-3} \cmidrule(lr){4-5} \cmidrule(lr){6-7} 
Country & F & M & F & M & F & M \\
\midrule
Austria     & \textBF{1.0222} & \textBF{0.8532} & 1.2664 & 0.9903 & 1.4321 & 1.2379  \\ 
Belgium     & \textBF{1.0125} & \textBF{0.8491} & 1.2066 & 1.0017 & 1.4340 & 1.2243 \\ 
Czech       & \textBF{0.8057} & \textBF{0.8487} & 1.0327 & 0.9047 & 1.1333 & 1.1743 \\ 
Denmark     & \textBF{1.3153} & \textBF{1.1100} & 1.6290 & 1.2535 & 1.8411 & 1.5491 \\ 
Estonia     & \textBF{1.5692} & \textBF{1.2700} & 1.8948 & 1.6387 & 1.8776 & 1.7259 \\ 
Finland     & \textBF{1.2332} & \textBF{1.0829} & 1.4799 & 1.1800 & 1.5812 & 1.1945 \\ 
France      & \textBF{0.5193} & \textBF{0.4959} & 0.6084 & 0.5143 & 0.9942 & 0.9589 \\ 
Hungary     & \textBF{1.1081} & \textBF{1.2734} & 1.3699 & 1.6906 & 1.6848 & 2.0493  \\ 
Iceland     & \textBF{1.7801} & \textBF{1.9743} & 1.8074 & 2.0974 & 1.9474 & 2.1992 \\ 
Ireland     & \textBF{1.3275} & \textBF{1.0769} & 1.5350 & 1.2967 & 1.7415 & 1.5085 \\ 
Italy       & 0.6039 & 0.5598 & \textBF{0.5811} & \textBF{0.5537} & 0.9152 & 0.8334 \\ 
Japan       & \textBF{0.5504} & \textBF{0.5068} & 0.5883 & 0.5644 & 0.8506 & 0.8603 \\ 
Latvia      & \textBF{2.0929} & \textBF{1.3091} & 2.6729 & 1.6556 & 2.7725 & 1.8484 \\ 
Lithuania   & \textBF{1.1972} & \textBF{1.6432} & 1.5601 & 2.0245 & 1.6413 & 2.2152 \\ 
Luxemburg   & \textBF{1.5977} & \textBF{1.8470} & 1.7749 & 1.9408 & 1.7865 & 2.0165 \\ 
Netherlands & \textBF{0.8442} & \textBF{0.7229} & 1.0810 & 0.9137 & 1.3713 & 1.1944 \\ 
New Zealand & \textBF{1.0994} & \textBF{0.9682} & 1.2688 & 1.1654 & 1.4546 & 1.3561 \\ 
Norway      & \textBF{1.2384} & \textBF{0.9516} & 1.5097 & 1.1664 & 1.6277 & 1.2927 \\ 
Poland      & 0.6349 & 0.6588 & \textBF{0.6279} & \textBF{0.5800} & 0.8487 & 0.9788 \\ 
Spain       & \textBF{0.6833} & 4.2268 & 0.7108 & \textBF{0.8513} & 1.0697 & 8.0521 \\ 
Sweden      & \textBF{0.9078} & \textBF{0.9066} & 1.1516 & 1.0646 & 1.2940 & 1.3398 \\ 
Switzerland & \textBF{1.2758} & 1.2596 & 1.5985 & \textBF{1.2532} & 1.8206 & 1.8009 \\ 
UK          & \textBF{0.5125} & \textBF{0.5604} & 0.5407 & 0.5748 & 0.9693 & 1.0729 \\ 
USA         & \textBF{0.4941} & \textBF{0.4785} & 0.5939 & 0.5278 & 1.1482 & 1.1103 \\
\cmidrule{2-7}
Mean        & \textBF{1.0594} & 1.1431 & 1.2538 & \textBF{1.1418} & 1.4682 & 1.6997 \\ 
\bottomrule
\end{longtable}
\end{center}

\vspace{-.5in}

The interior trimming and AIC methods generally yield the best interval prediction accuracy, as measured by the mean interval score. The SHAP-based ensemble prediction interval construction method outperforms the simple average method in most countries, with a significant reduction in the mean interval scores observed for both genders in Denmark, France, Hungary, Sweden, the USA, and for males in Belgium and Spain. Although the SHAP-based method beats the AIC-based ensemble method only for female mortality in Estonia, it produces smaller interval scores than the MSE-based method in France, Hungary, Switzerland, and the USA.

Truncating low-contributing models based on SHAP estimates improves interval forecast accuracy for Belgium, France, Spain, and the USA. Specifically, after excluding normalized SHAP values below $0.1$, the SHAP-based ensemble interval has a lower mean interval score than the interior trimming method. This is because the Renshaw-Haberman model in Table~\ref{tab:models_1} and M6, M7, M8 methods in Table~\ref{tab:models_2} produce inferior interval forecasts than the other methods for male mortality rates in Spain.

\section{Conclusion}\label{sec:7}

We propose a SHAP-based ensemble approach for forecasting mortality across multiple countries. By dynamically weighting multiple forecasting methods according to their marginal contributions, the ensemble ties together the complementary strengths of different models, whether from the Renshaw-Haberman, Cairns-Blake-Dowd, Lee-Carter, or functional time series models, thereby enhancing forecast accuracy compared to relying on any single model. These results show that different mortality models capture distinct aspects of mortality data across forecast horizons. Combining them in ensemble approaches can exploit their strengths while mitigating the drawbacks inherent in each model.

Furthermore, empirical analysis of data from 24 OECD countries suggests that SHAP-based ensembles outperform simpler combination schemes (e.g., simple model averaging or AIC-based weights) in terms of MSE and MAE. An additional advantage of this framework is its interpretability: the SHAP values indicate which models contribute most to overall forecast accuracy across different age groups, forecast horizons, and genders. 

Moreover, introducing a threshold on the SHAP values to exclude models with negligible or undermining contributions effectively reduces forecast variance without compromising overall accuracy. We also find that no single mortality model consistently stands out across all countries or for both genders, suggesting inherent model risk in this domain. The SHAP ensemble mitigates this issue by adaptively selecting and weighting models, accommodating variations in demographic patterns and data availability.

Future research could extend the proposed method to include a broader range of mortality models, investigate adaptive threshold choices, or explore integration with meta-learning techniques (e.g., stacked regression), further advancing the robustness and interpretability of mortality forecasting. Different countries may share a similar mortality profile, and clustering sets of homogeneous populations may improve the forecast accuracy. Since cluster membership impacts the computation of SHAP values, determining the optimal number of clusters using SHAP values remains an unexplored research question. Incorporating partial pooling of OECD country mortality rates into the construction of the SHAP ensemble could be a promising direction for future research.

\section*{Acknowledgment}
\vspace{-15pt}

The authors thank the editor, associate editor and three reviewers for their insightful comments and suggestions. The authors are grateful for the comments from the participants of the Gompertz 200 conference in Amsterdam, 2025. Han Lin Shang thanks financial support from an Australian Research Council Future Fellowship (FT240100338).

\section*{Funding}
\vspace{-15pt}

This research did not receive any specific grant from funding agencies in the public, commercial, or not-for-profit sectors.

\section*{Authors’ Contributions}
\vspace{-15pt}

All the authors contributed equally to the manuscript.

\section*{Conflict of Interest}
\vspace{-15pt}

The authors declare that they have no conflict of interest.

\newpage

\appendix

\newpage
\section{15 models and their constraints}\label{sec:Appendix_B}
\setcounter{table}{0}

\begin{table}[!htb]
\centering
\caption{The Renshaw-Haberman family models}\label{tab:models_1}
\renewcommand{\arraystretch}{1.4} 
\resizebox{\textwidth}{!}{%
\begin{tabular}{@{}llll@{}}
\toprule
\textbf{Label} & \textbf{Name} & \textbf{Model} & \textbf{Constraints} \\
\midrule
M1 & Lee-Carter (LC) - Poisson error
   & $\ln m_{x,t} = \alpha_x + \beta_x^{(1)} \kappa_t^{(1)}$
   & $\sum_t \kappa_t^{(1)} = 0,\; \sum_x \beta_x^{(1)} = 1$ \\
M2 & Renshaw-Haberman
   & $\ln m_{x,t} = \alpha_x + \beta_x^{(1)} \kappa_t^{(1)} + \beta_x^{(0)} \gamma_{t-x}$
   & $\sum_t \kappa_t^{(1)} = 0,\; \sum_x \beta_x^{(1)} = 1,\; \sum_x \beta_x^{(0)} = 1$ \\
M3 & Age-Period-Cohort
   & $\ln m_{x,t} = \alpha_x + \kappa_t^{(1)} + \gamma_{t-x}$
   & $\sum_t \kappa_t^{(1)} = 0,\; \sum_c \gamma_c = 0,\; \sum_c c\,\gamma_c = 0$ \\
\bottomrule
\end{tabular}
}
\end{table}

\begin{table}[!htb]
\centering
\caption{The Cairns-Blake-Dowd (CBD) family models}
\label{tab:models_2}
\renewcommand{\arraystretch}{1.4}
\resizebox{\textwidth}{!}{%
\begin{tabular}{@{}p{1cm}p{1cm}p{10cm}p{8cm}@{}}
\toprule
\textbf{Label} & \textbf{Name} & \textbf{Model} & \textbf{Constraints} \\
\midrule
M4 & CBD
   & $\ln m_{x,t} = \kappa_t^{(1)} + (x - \bar{x}) \kappa_t^{(2)}$
   & No constraints needed \\
M5 & M6
   & $\ln m_{x,t} = \kappa_t^{(1)} + (x - \bar{x})\,\kappa_t^{(2)} + (x_c - x)\,\gamma_{t-x}$
   & Similar to M7, but without quadratic term \\
M6 & M7 
   & $\ln m_{x,t} = \kappa_t^{(1)} + (x - \bar{x})\,\kappa_t^{(2)} + ((x - \bar{x})^2 - \sigma^2_x)\,\kappa_t^{(3)} + \gamma_{t-x}$
   & $\sum_t \kappa_t^{(1)} = 0$,\; $\sum_t \kappa_t^{(2)} = 0$,\; $\sum_t \kappa_t^{(3)} = 0$,\; $\sum_c \gamma_c = 0$,\; $\sum_c c^2\,\gamma_c = 0$ \\
M7 & M8
   & $\ln m_{x,t} = \alpha_x + \kappa_t^{(1)} + (x - \bar{x})\,\kappa_t^{(2)} + (x - \bar{x})^+\kappa_t^{(3)} + \gamma_{t-x}$
   & Similar to Plat Model \\
M8 & Plat 
   & $\ln m_{x,t} = \alpha_x + \kappa_t^{(1)} + (x - \bar{x})\,\kappa_t^{(2)} + (x - \bar{x})^+\kappa_t^{(3)} + \gamma_{t-x}$
   & $\sum_t \kappa_t^{(1)} = 0$,\; $\sum_t \kappa_t^{(2)} = 0$,\; $\sum_t \kappa_t^{(3)} = 0$,\; $\sum_c \gamma_c = 0$,\; $\sum_c c^2\,\gamma_c = 0$ \\
\bottomrule
\end{tabular}
}
\end{table}

\begin{table}[!htb]
\centering
\caption{The LC model and its variants}
\label{tab:models_3}
\renewcommand{\arraystretch}{1.4}
\resizebox{\textwidth}{!}{%
\begin{tabular}{@{}@{}p{1cm}p{7cm}p{6cm}p{8cm}@{}}
\toprule
\textbf{Label} & \textbf{Name} & \textbf{Model} & \textbf{Constraints} \\
\midrule
M9  & LC - Gaussian error
    & $\ln m_{x,t} = \alpha_x + \beta_x^{(1)} \kappa_t^{(1)} + \varepsilon_{x,t}$
    & $\sum_t \kappa_t^{(1)} = 0,\; \sum_x \beta_x^{(1)} = 1$ \\
M10 & BMS
    & $\ln m_{x,t} = \alpha_x + \beta_x^{(1)} \kappa_t^{(1)} + \beta_x^{(2)} \kappa_t^{(2)}$
    & $\sum_t \kappa_t^{(1)} = 0,\; \sum_x \beta_x^{(1)} = 1,\; \sum_x \beta_x^{(2)} = 1$ \\
M11 & LC - adjustment of life expectancy
    & $\ln m_{x,t} = \alpha_x + \beta_x^{(1)} \tilde{\kappa}_t^{(1)}$
    & $\sum_t \tilde{\kappa}_t^{(1)} = 0,\; \sum_x \beta_x^{(1)} = 1$ \\
M12 & LC - no adjustment to the score
    & $\ln m_{x,t} = \alpha_x + \beta_x^{(1)} \kappa_t^{(1)}$
    & No ex-post calibration constraints \\
\bottomrule
\end{tabular}
}
\end{table}

\begin{table}[!htb]
\centering
\caption{The Functional Time Series models}\label{tab:models_4}
\renewcommand{\arraystretch}{1.4}
\resizebox{\textwidth}{!}{%
\begin{tabular}{@{}p{1cm}p{3cm}p{9cm}p{8cm}@{}}
\toprule
\textbf{Label} & \textbf{Name} & \textbf{Model} & \textbf{Constraints} \\
\midrule
M13 & FDM
    & $m_{x,t} = \mu(x) + \sum_{k=1}^{K} \beta_{t,k} \,\phi_k(x) + \varepsilon_{x,t}$
    & Constraints on the first $K$ main components \\
M14 & Robust FDM
    & $m_{x,t} = \mu(x) + \sum_{k=1}^{K} \beta_{t,k} \,\phi_k(x) + \varepsilon_{x,t}$
    & Similar to Functional Data Model, but with robustness against outliers \\
M15 & Product-Ratio
    & $p_t(x) = \sqrt{m_M(x,t)\,m_F(x,t)},\quad r_t(x) = \sqrt{\frac{m_M(x,t)}{m_F(x,t)}}$
    & No specific constraints \\
\bottomrule
\end{tabular}
}
\end{table}

\section{MSE Decomposition: SHAP Weights vs.\ Equal Weights}\label{sec:app}

Previous work highlights how bias, variance, and diversity interact across horizons. \citet{taieb2015bias} show that multi-step strategies exhibit horizon-specific bias/variance behaviors; \citet{wood2023unified} emphasize that equal weighting tends to preserve diversity, which can be an effective variance-reduction device at short horizons. In our setting, equal weighting may therefore be competitive in the short run, while performance-weighted schemes (including SHAP) may gain ground as the horizon increases and conditional bias reduction becomes relatively more important.

Let $\mathcal{G}_h$ denote the out-of-sample evaluation set for horizon $h$, i.e., the collection of cells $j$ (e.g., age $\times$ cohort $\times$ sex) on which we compute losses. For each $j\in\mathcal{G}_h$, let $f_{h,j}\in\mathbb{R}^N$ be the (centered) vector of base forecasts; we form the combined forecast cell-by-cell as
\begin{equation*}
\widehat y_{h,j}=w'\,f_{h,j}.
\end{equation*}
We evaluate accuracy via the MSE,
\[
\mathrm{MSE}_h \;=\; \mathbb{E}\!\left[\big(y_{h,j}-\widehat y_{h,j}\big)^2\right]
\;=\; \underbrace{\big(\mathbb{E}[\widehat y_{h,j}]-y_{h,j}\big)^2}_{\mathrm{Bias}_h^2}
\;+\; \underbrace{\mathrm{Var}(\widehat y_{h,j})}_{\mathrm{Var}_h}
\;+\; \sigma^2,
\]
where $\sigma^2$ denotes irreducible noise. In our notation, $f_{h,j}=(\widehat{m}^{(1)}_{x_j,g_j,t_j+h},\ldots,\widehat{m}^{(N)}_{x_j,g_j,t_j+h})'$.

Throughout, variables are centred within each fold (equivalently, a fold-specific intercept is included), so that level bias is absorbed by the intercept while weights reflect the second-moment structure \citep{ElliottTimmermann2004}.

To link this decomposition to the combination literature, consider $\Sigma_{f,h}=\mathrm{Var}(f_{h,j})$ for the $N\times N$ within-fold covariance matrix of base forecasts at horizon $h$ (assumed positive definite, i.e., no exact collinearity).

\begin{proposition}[Conditional dominance]\label{prop:cond}
Under squared-error loss, and treating the combination weights as given, the conditional MSE at horizon $h$ is minimized by the regression solution $w_h^\star$ (\citealp{ElliottTimmermann2004}). For any alternative weights $w$,
\[
\mathrm{MSE}_h^{\mathrm{cond}}(w)
\;=\; \sigma_h^2 \;+\; (w-w_h^\star)'\,\Sigma_{f,h}\,(w-w_h^\star),
\]
where
\[
\mathrm{MSE}_h^{\mathrm{cond}}(w)\;=\;\mathbb{E}\!\big[(y_{h,j}-w' f_{h,j})^2\,\big|\,w\big]
\quad\text{and}\quad
\sigma_h^2=\mathrm{MSE}_h^{\mathrm{cond}}(w_h^\star).
\]
Hence $\mathrm{MSE}_h^{\mathrm{cond}}(w_h^\star)\le \mathrm{MSE}_h^{\mathrm{cond}}(w)$, with equality iff $w=w_h^\star$. If the SHAP-based weighting asymptotically recovers $w_h^\star$, it weakly dominates equal weights $w_{\mathrm{eq}}$ in conditional MSE.
\end{proposition}

\begin{remark}
\label{rem:finite}
Let $\widehat w_h$ be estimated weights and write $\Delta_h=\widehat w_h-w_h^\star$. The \emph{unconditional} MSE satisfies
\[
\mathbb{E}\!\left[\mathrm{MSE}_h(\widehat w_h)\right]
\;=\; \sigma_h^2
\;+\; \underbrace{\mathrm{tr}\!\left(\Sigma_{f,h}\,\mathrm{Var}(\widehat w_h)\right)}_{\text{estimation variance}}
\;+\; \underbrace{(\mathbb{E}[\widehat w_h]-w_h^\star)'\,\Sigma_{f,h}\,(\mathbb{E}[\widehat w_h]-w_h^\star)}_{\text{weight bias}}.
\]
For fixed schemes such as equal weights,
\[
\mathrm{MSE}_h^{\mathrm{cond}}(w_{\mathrm{eq}})=\sigma_h^2+(w_{\mathrm{eq}}-w_h^\star)'\Sigma_{f,h}(w_{\mathrm{eq}}-w_h^\star).
\]
Therefore, SHAP-based weights improve out-of-sample performance only if the conditional reduction relative to $w_{\mathrm{eq}}$ exceeds the estimation-variance (and any weight-bias) penalty. This bias–variance trade-off underlies the forecast combination puzzle \citep{ElliottTimmermann2004,ElliottTimmermann2005,SmithWallis2009,GenreKennyMeylerTimmermann2013}.
\end{remark}

Our conditional statement aligns with \citet{ElliottTimmermann2004}: under squared-error loss, the weights that minimize conditional MSE coincide with the regression solution, while centering (or a fold-specific intercept) absorbs level bias so that weights reflect second moments. When relative performance varies over time, \citet{ElliottTimmermann2005} show why constant least-squares weights may underperform and why adaptive or regime-dependent combinations can dominate, consistent with Shapley-based weights gaining traction as horizons lengthen.

The qualification in Remark~\ref{rem:finite} connects directly to the combination puzzle: sampling variability in estimated weights can offset conditional gains, making simple averages competitive with short samples, high collinearity, or unstable covariances \citep{SmithWallis2009}. Large comparative studies similarly find that equal weights are hard to beat robustly \citep{GenreKennyMeylerTimmermann2013}. In our setting, SHAP-based weights improve conditional fit when they approximate the regression solution, but their finite-sample edge depends on the estimation-variance penalty.

\subsection{MSE decomposition for SHAP weights}

For an ensemble with SHAP weights,
\[
\widehat{m}^{(c)}_{h,\text{Shap}}(x,g,t+h)=\sum_{i=1}^{N^\ast}\widehat{\omega}_{i,\text{Shap}}\,\widehat{m}^{(i)}_{x,g,t+h},
\]
the MSE decomposes as
\[
\mathrm{MSE}_{h,\text{Shap}}
=\mathrm{Bias}_{h,\text{Shap}}^2+\mathrm{Var}_{h,\text{Shap}}+\sigma^2,
\]
with
\begin{align*}
\mathrm{Bias}_{h,\text{Shap}}^2
&=\Big(\sum_{i=1}^{N^\ast}\widehat{\omega}_{i,\text{Shap}}\,\mathbb{E}\,[\widehat{m}^{(i)}_{x,g,t+h}] - y_{t+h}\Big)^2\\
\mathrm{Var}_{h,\text{Shap}}
&=\sum_{i}\widehat{\omega}_{i,\text{Shap}}^2\,\mathrm{Var}(\widehat{m}^{(i)})+\sum_{i\neq j}\widehat{\omega}_{i,\text{Shap}}\widehat{\omega}_{j,\text{Shap}}\,\mathrm{Cov}(\widehat{m}^{(i)},\widehat{m}^{(j)}).
\end{align*}

\subsection{MSE decomposition for equal weights}

For equal weights $\widehat{\omega}_i=1/N$,
\[
\mathrm{MSE}_{h,\text{equal}}
=\mathrm{Bias}_{h,\text{equal}}^2+\mathrm{Var}_{h,\text{equal}}+\sigma^2,
\]
with
\begin{align*}
\mathrm{Bias}_{h,\text{equal}}^2
&=\Big(\tfrac{1}{N}\sum_{i=1}^{N}\mathbb{E}\,[\widehat{m}^{(i)}_{x,g,t+h}] - y_{t+h}\Big)^2, \\
\mathrm{Var}_{h,\text{equal}}
&=\tfrac{1}{N^2}\sum_{i}\mathrm{Var}(\widehat{m}^{(i)})+\tfrac{1}{N^2}\sum_{i\neq j}\mathrm{Cov}(\widehat{m}^{(i)},\widehat{m}^{(j)}).
\end{align*}

Taken together with Proposition~\ref{prop:cond} and Remark~\ref{rem:finite}, these decompositions imply that the relative performance of SHAP versus equal weights is horizon-dependent through the bias–variance trade-off and the sampling variability of the estimated weights: at short horizons, diversity-driven variance reduction can make equal weights competitive; as the horizon increases, SHAP-based weighting may better approximate the regression solution and reduce conditional MSE.

\subsection{MSE decomposition with SHAP-based thresholds}

To improve stability, we introduce a threshold $\alpha$ that excludes models with negligible contributions:
\[
S(\alpha)=\big\{i:\,\widetilde{\phi}_i(\nu;g)>\alpha\big\},\qquad
\widehat{m}^{(c)}_{h,\text{Shap}\,\alpha}=\sum_{i\in S}\widehat{\omega}_{i,\text{Shap}\,\alpha}\,\widehat{m}^{(i)}_{x,g,t+h}.
\]

Because $S(\alpha)$ is nested in $\alpha$, increasing $\alpha$ weakly reduces the remaining set. Substituting the truncated weights yields:
\begin{align}
\mathrm{Bias}_{h,\text{Shap}\,\alpha}^2
&= \Big(\sum_{i\in S(\alpha)} \widehat{\omega}_{i,\text{Shap}\,\alpha}\,
         \mathbb{E}[\widehat{m}^{(i)}_{x,g,t+h}] - y_{t+h}\Big)^2,
\label{eq:bias_shap_alpha}\\[4pt]
\mathrm{Var}_{h,\text{Shap}\,\alpha}
&= \sum_{i\in S(\alpha)} \widehat{\omega}_{i,\text{Shap}\,\alpha}^2\,
      \mathrm{Var}(\widehat{m}^{(i)})\;+\!
      \sum_{\substack{i\neq j\\ i,j\in S(\alpha)}}\!
      \widehat{\omega}_{i,\text{Shap}\,\alpha}\widehat{\omega}_{j,\text{Shap}\,\alpha}\,
      \mathrm{Cov}(\widehat{m}^{(i)},\widehat{m}^{(j)}).
\label{eq:var_shap_alpha}
\end{align}

Filtering low-contribution models generally reduces variance -- especially when excluded models are unstable -- while preserving bias. However, overly aggressive thresholds can reduce diversity, potentially increasing bias and, if negative covariances are removed, even inflating variance. In practice, the threshold acts as a form of shrinkage: moderate values improve short-horizon robustness, whereas excessive filtering can harm long-horizon performance by limiting diversity.

\newpage
\bibliographystyle{agsm}
\bibliography{ensemble}

\end{document}